\documentclass[journal]{IEEEtran}
\usepackage{cite}

\usepackage{subfig}
\usepackage{pstricks}
\usepackage{pst-plot}
\usepackage{pst-node}
\usepackage{multido}
\usepackage{pstricks-add}
\usepackage{pst-eucl}
\usepackage{pst-intersect}
\usepackage{environ}

\usepackage{pst-optexp}

\ifCLASSINFOpdf
\usepackage[pdftex]{graphicx}

\else
  \usepackage[dvips]{graphicx}
\fi
\usepackage[cmex10]{amsmath}
\usepackage{amssymb}
\begin{document}
%
\title{Hidden probe attacks on ultralong fiber laser key distribution systems}
%
%
%

\author{Juan~Carlos~Garcia-Escartin 
        and~Pedro~Chamorro-Posada.
\thanks{Corresponding author: juagar@tel.uva.es. Universidad de Valladolid, Dpto. de Teor\'ia de la Se\~{n}al y Comunicaciones e Ingenier\'ia Telem\'atica, Paseo Bel\'en n$^{o}$ 15, 47011 Valladolid, Spain.}
}

%
%

\markboth{}%
{}
%



\maketitle

\begin{abstract}
In ultralong fiber laser key distribution, two sides use standard optical equipment to create kilometer long fiber lasers in a communication link to establish a secret key. Its security rests on the assumption that any attacker would need much more sophisticated equipment and techniques than those of the legitimate user in order to discover the generated key. We present a challenge to that assumption with a hidden probe attack in which the eavesdropper hides a weak signal in the unavoidable noise floor that appears in the laser during amplification and probes with it the configuration of one or both communication parties. We comment how this attack can compromise different proposals for ultralong laser key distribution and propose possible countermeasures.
\end{abstract}


%
\IEEEpeerreviewmaketitle

\section{Key distribution}
%
%
%
%
\IEEEPARstart{K}{ey} distribution is a fundamental problem in modern communication networks. When two sides, Alice and Bob, want to establish a a private communication channel, they usually turn to symmetric cryptosystems, such as AES \cite{AES}. For convenience and security, most symmetric cryptosystems follow Kerckhoffs’s principle \cite{Ker83}: the users assume all the details of the implementation are known and the security rests on a shared secret key known only to Alice and Bob. That way, if a key is stolen, they can immediately switch to a new one without changing anything else. 

A key distribution method gives a way to generate these secret keys and communicate them securely to both parties. Usually, there is first a slow key distribution procedure that establishes a new key for each communication session between Alice and Bob. Then, the generated session key is used in a fast symmetric cryptosystem to send the data. These session keys are relatively short, usually no more than a few hundred of bits long, which makes the initial slower distribution acceptable.

In current computer networks, key distribution is enabled by asymmetric, or public key, cryptography. Public key cryptography is based on the assumed asymmetry of certain mathematical problems. One example is the RSA cryptosystem \cite{RSA78}, in which privacy is protected by the fact that, while multiplying two large primes, $p$ and $q$, is straightforward, there are no known algorithms for fast factorization of a number $N=pq$ except for Shor's algorithm \cite{Sho97}, which would require a full scale quantum computer, which seems unlikely to be built in the near future. 

Since the appearance of Quantum Key Distribution, QKD \cite{BB84,Eke91}, there has been a growing interest in key distribution systems where the protection comes from physical principles. The main problem in physical key distribution is finding an adequate channel which gives more information to Alice and Bob than it leaks to Eve. In quantum key distribution, the laws of Quantum Mechanics guarantee that eavesdroppers are detected. The protection comes from the imposibility of measuring a quantum system without disturbing its state. Alice and Bob can find out if there is an eavesdropper and, after public discussion with a privacy amplification protocol that uses a small shared key \cite{BBR88,BBC95}, they can produce long keys known only to the two of them. For quantum communication channels there exist rigurous proofs of the theoretical security of the scheme under different assumptions \cite{SP00,GLL04,RGK05,LC99,Lut00,May01}. However, practical details of the implementation can lead to deviations from the ideal system which can be exploited in different attacks \cite{FQT07,QFL07,ZFQ08,XQL10,LWW10a,LWW10b,LAM11,LSM11,WLW11,SRK15} and it is essential to be careful when building a practical system.  

Spurred by the success of QKD, there have appeared different proposals for physical key distribution with varying degrees of protection using optical or electrical systems \cite{Kis06,MGK08,Liu09,SY06b,Liu09b,Liu09c,Liu10}. In this paper, we discuss key distribution systems with ultralong lasers, following the proposal of Scheuer and Yariv of an optical system that builds on long distance fiber links to create a long fiber laser cavity between Alice and Bob \cite{SY06}.

The basic concept behind this and most alternative proposals for physical key distribution is the ``keyless cryptography'' protocol of Alpern and Schneider \cite{AS83}. The protocol gives a way for Alice and Bob to agree on a secret key using a public anonymous channel. Both Alice and Bob can broadcast a list of bits for a random key they have generated, announcing them as their common key proposal. Anyone can see the proposed values $k_{AB}[i]$ for the $i$th bit of the key from both Alice and Bob, but, if the channel is anonymous, only Alice and Bob know which bit belongs to each one of them. They can check all the bits in the sequence and, if the bits are equal, they discard them. If they are different they can choose, for instance, Alice's bit as the next key bit. 

The lack of an adequate public anonymous channel has stopped the practical implementation of the ``keyless cryptography'' protocol. The proposals for physical key distribution systems try to solve this problem. 

In ultralong laser key distribution, instead of aiming for a complete proof of security, like in QKD, the purpose is to introduce a physical asymmetry, much like the computational asymmetry in public key algorithms. An analogy in terms of an everyday physical security device would be a combination safe. It is not physically impossible for a thief to open it, but, if properly designed, it introduces a degree of complication that deters most attempts while it imposes a light burden on the user.

In an ultralong fiber laser key distribution system, the legitimate users would only need to set up a relatively simple physical system and perform standard measurements but an eavesdropper would require more sophisticated equipment and methods. We suggest a possible attack that would lower that barrier for the attacker, who would only need to hide probe pulses in the floor noise of the fiber laser. To our knowledge, this is the first detailed analysis of an active attacks against UFL key distribution systems and shows it should be considered as an important threat. 

In Section \ref{UFLKD}, we present the fundamental principles of ultralong fiber laser key distribution and discuss three alternative implementations. In Section \ref{ProbeAttack} we present the hidden probe attack using spread spectrum modulation and give a brief analysis of its possibilities and limitations. Section \ref{Results} gives the simulation results for different hidden probe attacks on the three ultralong fiber laser key distribution implementations under study. Finally, in Section \ref{Discussion}, we discuss the extent of this attack and propose potential countermeasures.

\section{Ultralong fiber laser key distribution}
\label{UFLKD}
Fiber lasers use optical fiber as the gain medium of a laser. Fiber lasers can have a cavity spanning for hundred of kilometers \cite{AEI06,TAB09,TBE10}. These ultralong fiber lasers, UFLs, can be built over existing fiber communication links. A fiber link with fiber mirrors at the ends gives a suitable laser cavity. For instance, we can use fiber Bragg gratings to reflect light to chose a resonating frequency from the allowed modes. The laser gain comes from a pump signal at another frequency and nonlinear conversion. The gain of the cavity can be distributed along the fiber or concentrated on small sections of the whole length. UFLs can use Raman amplification, Erbium-Doped Fiber Amplifiers, EDFAs, or Semiconductor Optical Amplifier, SOAs.  

In ultralong fiber laser key distribution, the final state in the fiber laser cavity is tuned from the ends, which host each communication party. Alice and Bob can change their setup to alter the cavity and determine the frequency of the lasing signal. Each one has two alternative configurations, 0 and 1, and there are four possible states for the cavity, 00, 01, 10 and 11, where the first bit describes the configuration chosen by Alice and the second bit the configuration chosen by Bob. This state is public and Eve can measure it at any time with negligible impact on the state. In a properly designed system, the stationary states for the configurations 01 and 10 will be extremely similar and Eve will not be able to tell them apart without a considerable effort. If Alice and Bob choose their configurations randomly and produce a new state every $T$ seconds, they can agree on a secret key by discarding all the time bins where their chosen bits were the same. All three, Alice, Bob and Eve, can monitor the channel and know when the configurations 01 and 10 appear, but only Alice and Bob have a simple way to tell apart 01 from 10. Alice knows her chosen bit, so Bob's bit must have chosen the opposite value. Bob can do a similar deduction. On average, half of the time bins there will be a 01 or 10 state, which gives one bit of secret key. Alice and Bob can simply use Alice's choice, which they both know, as their key bit.

The assymetry comes from the additional information Alice and Bob have. They both know their own choice of configuration and do not need to distinguish 01 and 10 states from measurements of the channel. Eve does not have that luxury and must try to establish the difference from channel measurements or by peering inside Alice's or Bob's setup. 

Eve's strategies can be grouped into active and passive attacks. In passive attacks, Eve just reads the channel, tapping the fiber as unobtrusively as possible. Her main challenge is analyzing the data to tell apart the states 01 and 10. She can look into transient states or study the spectrum \cite{ZSS08,EKH14}. In active attacks, Eve will introduce a signal to discover the setup on Alice's or Bob's side. The usual assumption is that any active attacks would show due to the gain in the laser \cite{SY06}. Any probes send to determine Alice's or Bob's state would produce a conspicuous change in the channel and the exchange could be aborted until no eavesdropping is detected. In the previous literature, active attacks have been put aside on those grounds. In the rest of the paper, we will show they are indeed an important concern. If Eve manages to hide probe in the floor noise that appears during amplification, the change in the channel is difficult to detect and sneak active attacks become feasible.

There are multiple experimental and theoretical results for UFL key distribution with different configurations \cite{BS09,KS10,EKH14}. We will evaluate our hidden probe attack against three main families of implementations: a basic setup following the original proposal of Scheuer and Yariv \cite{SY06}, which uses frequency selection, a length-based key distribution system \cite{TBK15} and a dark state implementation with lasing and non-lasing stationary states \cite{KS14}.

\subsection{Basic setup: selective frequency setup}
\begin{figure}[!ht]
\centering
\includegraphics[width=\columnwidth]{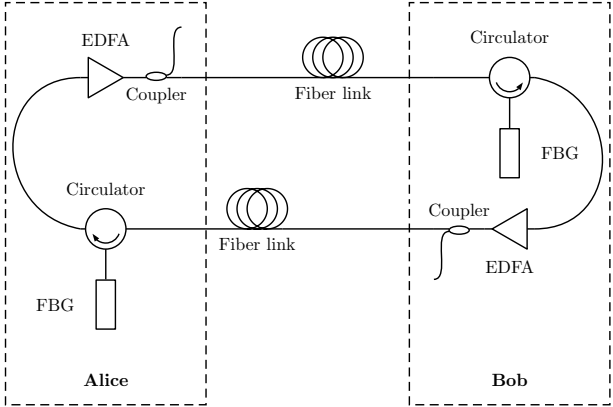}
\caption{UFL key distribution with frequency selection. Alice and Bob can change the state inside the fiber laser cavity by tuning the reflection peak of a Fiber Bragg Grating, FBG.}
\label{basic}
\end{figure}

Our reference system will be the original proposal of UFL key distribution of Scheuer and Yariv \cite{SY06} and its ring implementation \cite{ZSS08} shown in Fig. \ref{basic}. The gain of the fiber laser comes from two Erbium-Doped Fiber Amplifier, EDFAs, one at each side. The cavity is closed by two tunable mirrors. Alice and Bob each have a Fiber Bragg Grating, FBG, which can be tension-tuned to give peak reflectivity frequencies $f_0$ or $f_1$. If both mirrors have the same peak frequency, the fiber laser will resonate at that value $f_0$ or $f_1$. For the 01 and 10 configurations where Alice and Bob choose a different mirror, there will be a signal oscillating close to the average frequency $f_{c}=\frac{f_0+f_1}{2}$. 

The signals for the 01 and 10 configurations are difficult to distinguish, which gives the desired behaviour for the key distribution procedure we outlined in the previous Section. 

The system at each side is completed with a coupler that samples the resulting signal. This gives Alice and Bob the ability to read the channel to see which of the possible stationary states, 00, 01, 10 or 11, has been reached and to check for active attacks.

In our simulations, we will use the system as described in the experimental realization given in\cite{ZSS08}.

\subsection{Length variation setup}
\begin{figure}[!ht]
\centering
\includegraphics[width=\columnwidth]{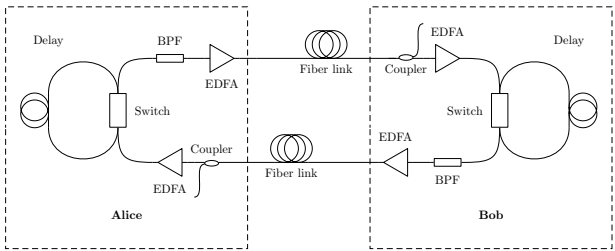}
\caption{UFL key distribution with length selection. Alice and Bob can change the state inside the fiber laser cavity by switching between a short and a long fiber loop.}
\label{length}
\end{figure}

An alternative way to alter the fiber cavity is changing its length. Figure \ref{length} shows the diagram of an experimental demonstration of UFL key distribution based on this principle \cite{TBK15}. Instead of selecting the mode by attenuating all the signals outside a certain frequency range, a variable delay loop changes the free spectral range of the fiber laser cavity, resulting in a frequency shift in the lasing signal. 

Alice and Bob have a one kilometer delay line they can selectively switch on or off. The 0 and 1 configurations on each side correspond to the short and long lines respectively. The total length of the fiber determines the lasing frequency, with kHz shifts for the chosen values in the experiment (a total of 50 km of fiber in the 00 configuration, which becomes 51 km for the 01 and 10 states and 52 km for the 11 setup). Again, while the states corresponding to 00 and 11 can be easily told apart, the 01 and 10 states are almost identical. 

The system is completed on each side with a bandpass filter, BPF, to keep within the frequency range of interest in the laser and monitoring couplers that allow the users to sample the channel and check for attacks.

\subsection{Dark state setup}
The third implementation we will consider is a dark state ultralong fiber laser key distribution system \cite{KS14}. The scheme is essentially the basic frequency selective setup of Fig. \ref{basic}, but adding an interferometer before the output of Alice and Bob that acts as a narrow bandpass filter so that only signals very close to $f_0$ or $f_1$ get through. In this case, there is only lasing for the 00 and 11 configurations and the 01 and 10 states correspond to the noise floor which makes state identification more difficult.

The term ``dark state'' refers to the feature that, now, the bits of the key are generated for a fibee without a laser oscillation (a dark fiber which only carries noise).

\section{Hidden probe attacks}
\label{ProbeAttack}
\begin{figure}[!ht]
\centering
\includegraphics[width=\columnwidth]{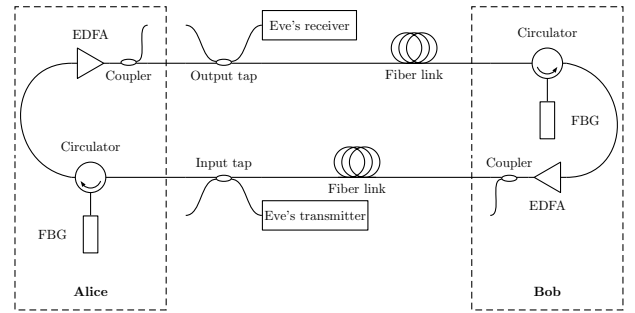}
\caption{Attack setup: Eve will tap the fiber close to Alice to learn her state. In the input to Alice, she will introduce a modulated signal from a transmitter and can, optionally, monitor the channel. In the output from Alice, Eve will measure the light in the fiber to recover the probe signal and analyze it to determine the configuration inside Alice. Eve can also use that signal to monitor the global state of the laser.}
\label{attack}
\end{figure}

In all the presented UFL key distribution systems, there appears a noise signal during the gain stage which allows Eve to hide a probe signal. Some systems even artificially inject more noise as a countermeasure to passive attacks where Eve measures the transient response of the system \cite{ZSS08}. In the active attack we propose, Eve introduces a signal below the noise floor to learn the configuration of one of the side's (Alice in our example). The general attack is described in Fig. \ref{attack} for the basic UFL key distribution system, but it would be the same for the rest of the setups. 

\begin{figure}[!t]
\centering
\subfloat[Eve's transmitter with spread spectrum modulation using a pseudorandom, PR, sequence, and a phase modulator PM.]{\includegraphics[width=\columnwidth]{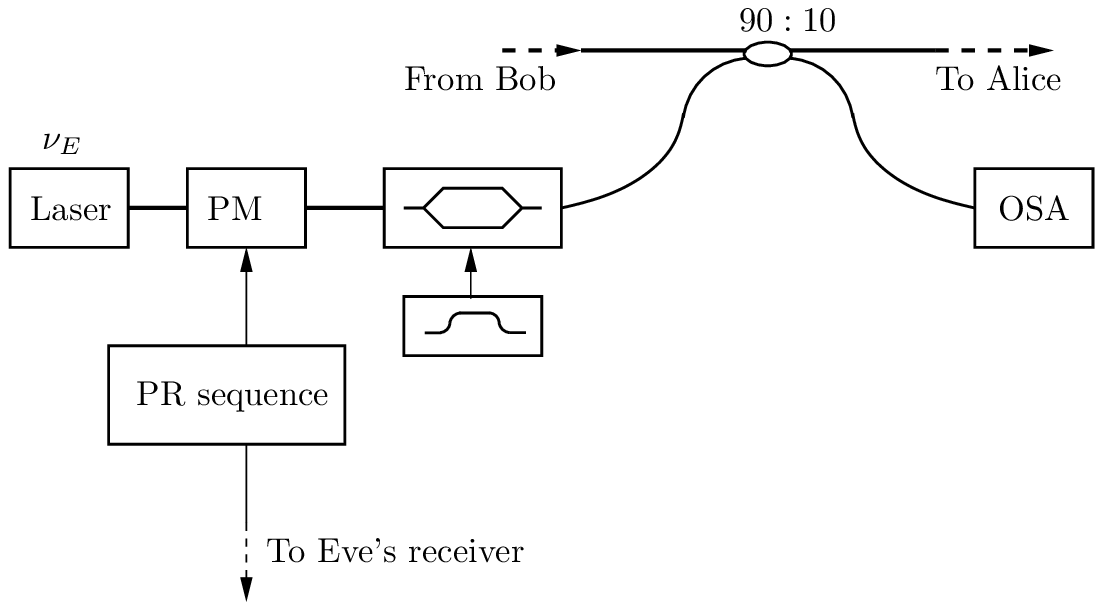}%
\label{ETx}}
\hfil
\subfloat[Eve's receiver with direct detection.]{\includegraphics[width=\columnwidth]{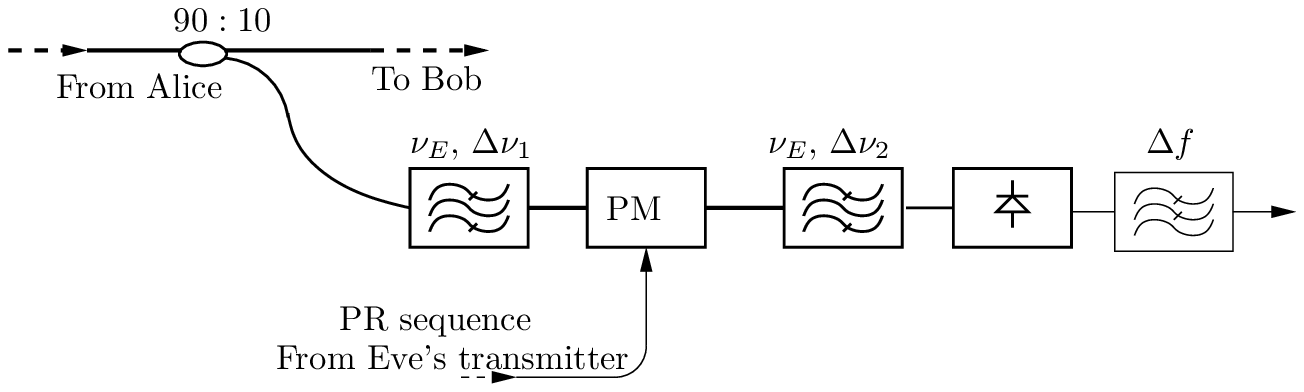}%
\label{ERxI}}
\hfil
\subfloat[Eve's receiver with coherent detection.]{\includegraphics[width=\columnwidth]{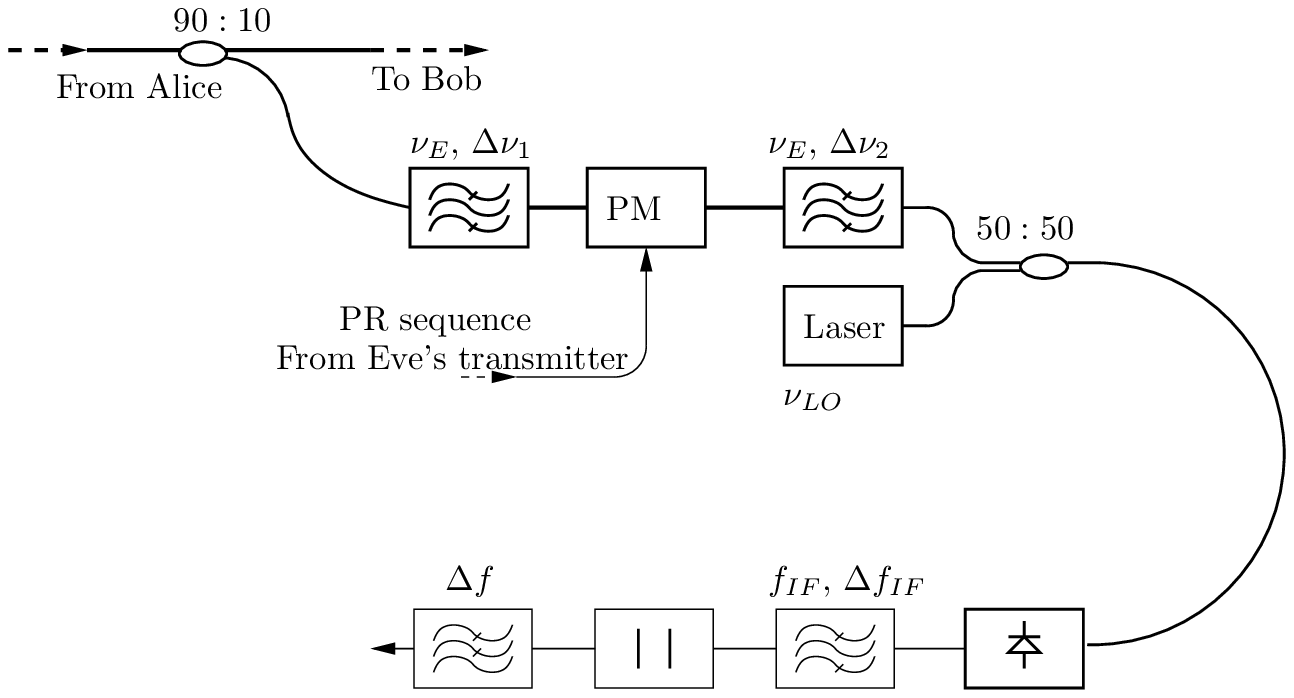}%
\label{ERxC}}
\caption{Eve's transmitter (\ref{ETx}) and receiver for incoherent (\ref{ERxI}) and coherent (\ref{ERxC}) detection.}
\end{figure}

Eve will send a spread spectrum signal extending through all the available bandwith. She will prepare a probe pulse and then modulate it with a phase modulator, PM, controlled by a pseudo-random code, the PR sequence \cite{Mut96}, so that the total power is stretched in the band of interest. This will reduce the peak power which allows to hide the signal in the noise of the fiber laser. The resulting signal will be modulated in time so that it will only be active for a probe time $T_p=1$ ms, smaller than the total bit time $T$. The limited time frame helps to avoid affecting the transients in the laser and makes sure the probe is only sent when there is an established state in the fiber. 

The probe signal is sent inside of Alice's setup. For the basic, frequency selective implementation, Eve will center her probe at one frequency, for instance $f_0$. If the mirror of Alice is in the 0 configuration, the receiver Eve has placed at Alice's output will detect the probe pulse. A matching demodulator using the same PR sequence permits to recover the original pulse with a processing gain with respect to the noise proportional to the spreading factor \cite{PSM82}, producing a clear signal. However, if Alice's mirror is centered around $f_1$, most of the probe signal will be destroyed and Eve will measure a noise-like signal. The result of her measurement gives her a way to determine the configuration inside Alice. In the dark state implementation the principle will be similar, with some technical differences (see Section \ref{DarkResults}).

For length coding, Eve can check for the autocorrelation of the probe signal with a delayed version of itself. The delay can be set to be equivalent to either the short or the long configuration. If there is a peak at the receiver, Eve will know she has chosen the same bit as Alice. If not, she knows Alice has chosen the other value. 

We assume Eve can synchronize her pulses either because she knows the internal configuration of Alice, except for her random choices, and can predict the total delay of the probe or because Eve has determined the delay from autocorrelation measures of previous probes before launching the definitive attack. 

The use of spread spectrum modulation to hide a signal has been tested before in optical fiber communications in optical steganography \cite{WPN06,WN06}. In optical steganography, the end users try to establish a covert channel so that an external eavesdropper does not know any communication is taking place \cite{WSM15}. Many proposals for optical steganography use spread spectrum modulation and related schemes to hide a communication channel in the Amplified Spontaneous Emission,or ASE, noise from the EDFAs \cite{HWG07,WP11,WWT13,WWS14}. Optical steganography also shows ways to use a covert channel for synchronization \cite{WFX10}, which could be useful if do not have an alternative way to establish a clock.

Before presenting the results of the simulation, we would like to examine the ultimate limitations of probe attacks. Shannon's noisy channel theorem \cite{Sha49a} tell us the maximum possible communication rate for analog communication is given by
\begin{equation}
C=\frac{B}{T}=W\log_2 (1+SNR)
\end{equation}
where $C$ is the number of bits per second (or $B$ bits in a time segment $T$) that can be reliably transmitted using a bandwith $W$ when our signal to noise ratio is $SNR$. 

Eve's hidden probes should pass unnoticed. The EDFA means strong active probes will stand out when measuring the channel. This limits the SNR of our probes, which should have a power below the noise floor of the EDFAs that will hide them. A brief informal calculation shows such an attack is plausible with weak probes. Taking as the bandwith the usual separation between $f_0$ and $f_1$, in the range of GHz, and a time per bit around 1 ms to allow for a stationary state to build up, even for a probe 30 dB below the noise floor, $SNR=10^{-3}$, we could send around $B=TW\log_2(1+SNR)\lesssim 1442$ bits if we had and appropriate codebook of random signals. Actually, what we have is two different channels where we can imagine Alice is modulating the data with two waveforms, one for each configuration. While the actual bit rate is likely to be smaller than this best case limit, this rapid calculation shows there is plenty of room for a probe attack.

Optical steganography has also proposals with matching gratings \cite{FP09} which remind of the basic UFL key distribution system and could suggest which waveforms would perform better in taking information out of Alice. 

\section{Simulation results}
\label{Results}
The simulations of the UFL dynamics have been based on the models of \cite{SY06,ZSS08}, which have been adapted for the three UFL configurations considered.  We assume fiber sections of $25$ km and a typical signalling time of $T=3$ ms and allow Eve for a probe time of $T_p\lesssim 1$ ms.  This permits Eve to listen to the laser channel and decide if a useful secret key transmission is taking place.  The simulation of Eve's channel probing is performed in the presence of the corresponding steady state cavity field.   

The schematic of Eve's transmitter is shown in Fig. \ref{ETx}.  Eve has to carefully adjust her laser output power (for instance, using a variable attenuator) in order to avoid the detection of her presence.  The CW optical signal is then phase modulated using a pseudorandom sequence and its amplitude shaped with a $T_p$ pulse.  We have assumed that Eve's laser has a $100$ kHz linewidth.  In all cases, the fiber attenuation has been set to $0.2$ dB/km and the fiber group index to $n_g=1.462$.

The phase modulators at Eve's receiver and transmitted are assumed to act synchronously.  Group delays both in Alice's setup and also in Eve's equipment are, therefore, most relevant and their effect is discussed in detail below.

Eve can perform either a direct or a heterodyne coherent detection on her receiver, as described in Figs. \ref{ERxI} and \ref{ERxC} respectively.  The phase modulation is a highly nonlinear operation that, besides the desired effect of recovering the temporal pulse shape of Eve's probing signal, will mix Eve's noise.  For this reason, bandpass optical filters are placed both before and after the phase modulator.  The filter specifications will be set by the particular UFL and Eve's system implementations. In the coherent detection receiver, for instance, the response of the optical filters will be determined by the value of the RF intermediate frequency corresponding to the beating of the Eve's lasers.  In the numerical simulations, where a limited overall system bandwidth is considered, these filters are very important to obtain a faithful representation of the modelled system and the specific filter parameters used obey mainly to numerical considerations.  We have used very steep supergaussian $m=4$ spectral amplitude responses with $\Delta\nu_1=2$ GHz and $\Delta\nu_2=200$ MHz. The bandwidth of the bandpass intermediate frequency filter in Fig. \ref{ERxC} used in the calculations is $\Delta f_{FI}=20$ MHz and the lowpass filter bandwith $\Delta f=10$ kHz.  For such a small filter bandwidth, thermal noise at the reciver can be made to remain below the signal level in those of cases commented below where the detection of Eve's signal is sucessful, and has been ignored in the numerical calculations.      

Eve's photodetector is assumed to have a responsivity of $0.8$ A/W and her local oscillator laser has a linewidth of $100$ kHz and $1$ mW of optical power.  An increase of the local oscillator power will have a direct impact on the receiver performance.

\subsection{Basic setup}

\begin{figure}[!ht]
\centering
\includegraphics[width=0.7\columnwidth]{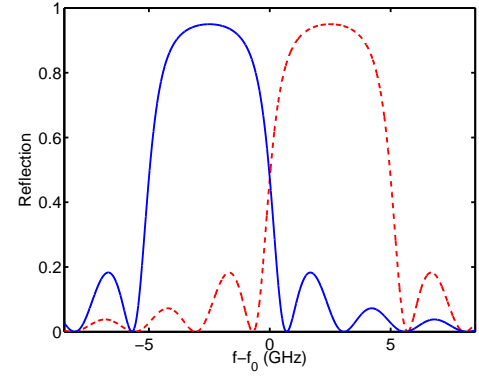}
\caption{Amplitude response of the gratings used both by Alice and Bob to set their respective state.}\label{filtros}
\end{figure}

The spectral responses of the gratings used by Alice and Bob to code their information have a relative shift of $5$ GHz and have identical FWHM spectral widths of $5$ GHz.  Their amplide responses are shown in Fig.\ref{filtros}.  The EDFAs are assumed to have a small signal gain of $G=20$ dB, a saturation power of $Psat=13$ dBm, and a noise figure of $F=4.5$ dB. The optical power measured by Alice at her output coupler port during effective key transmission is $P_{out} \simeq 19.5$ mW.

\begin{figure}[!ht]
\centering

\begin{tabular}{cc}
{ (a)}&{ (b)}\\
\includegraphics[width=0.45\columnwidth]{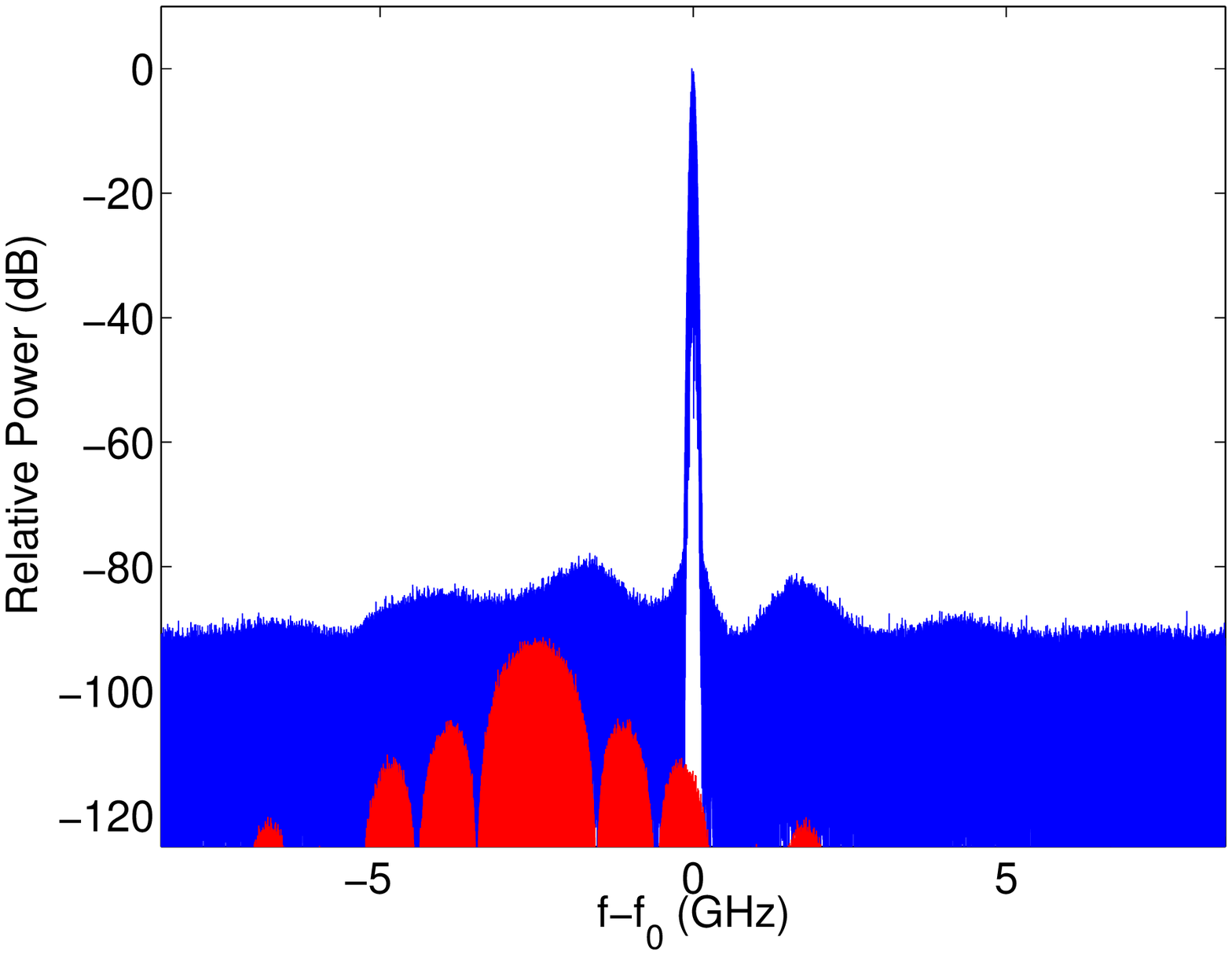}&\includegraphics[width=0.45\columnwidth]{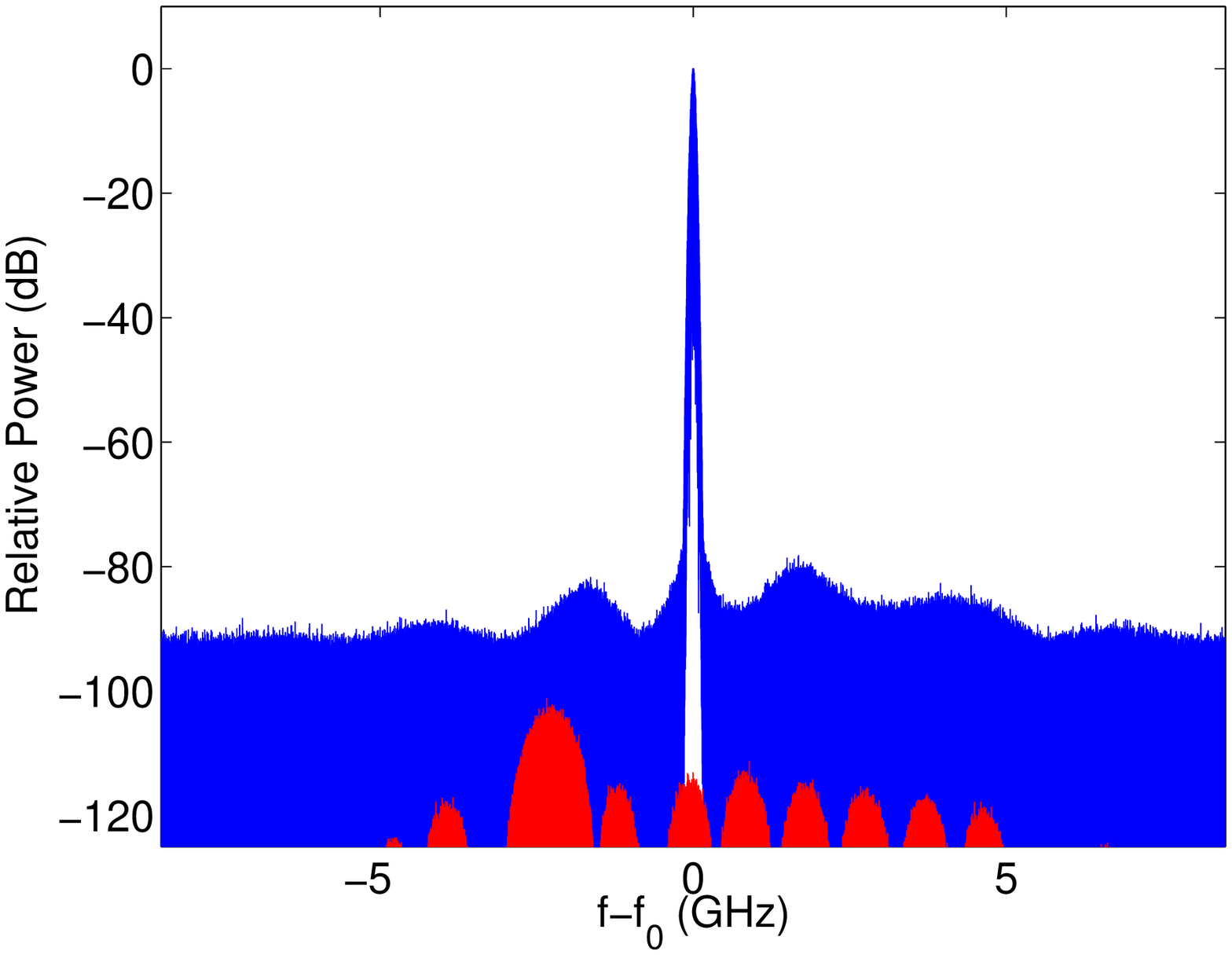}\\
{ (c)}&{ (d)}\\
\includegraphics[width=0.45\columnwidth]{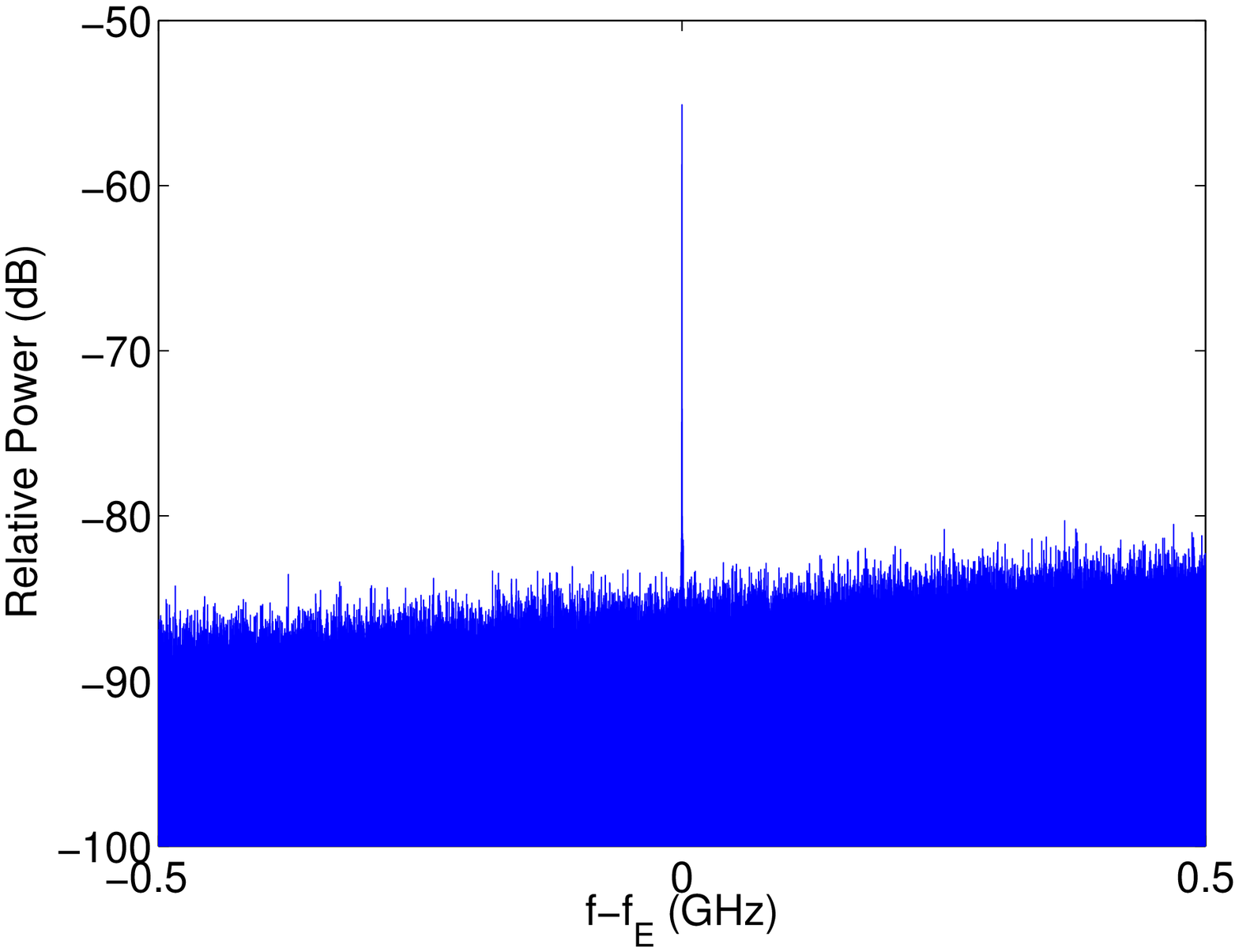}&\includegraphics[width=0.45\columnwidth]{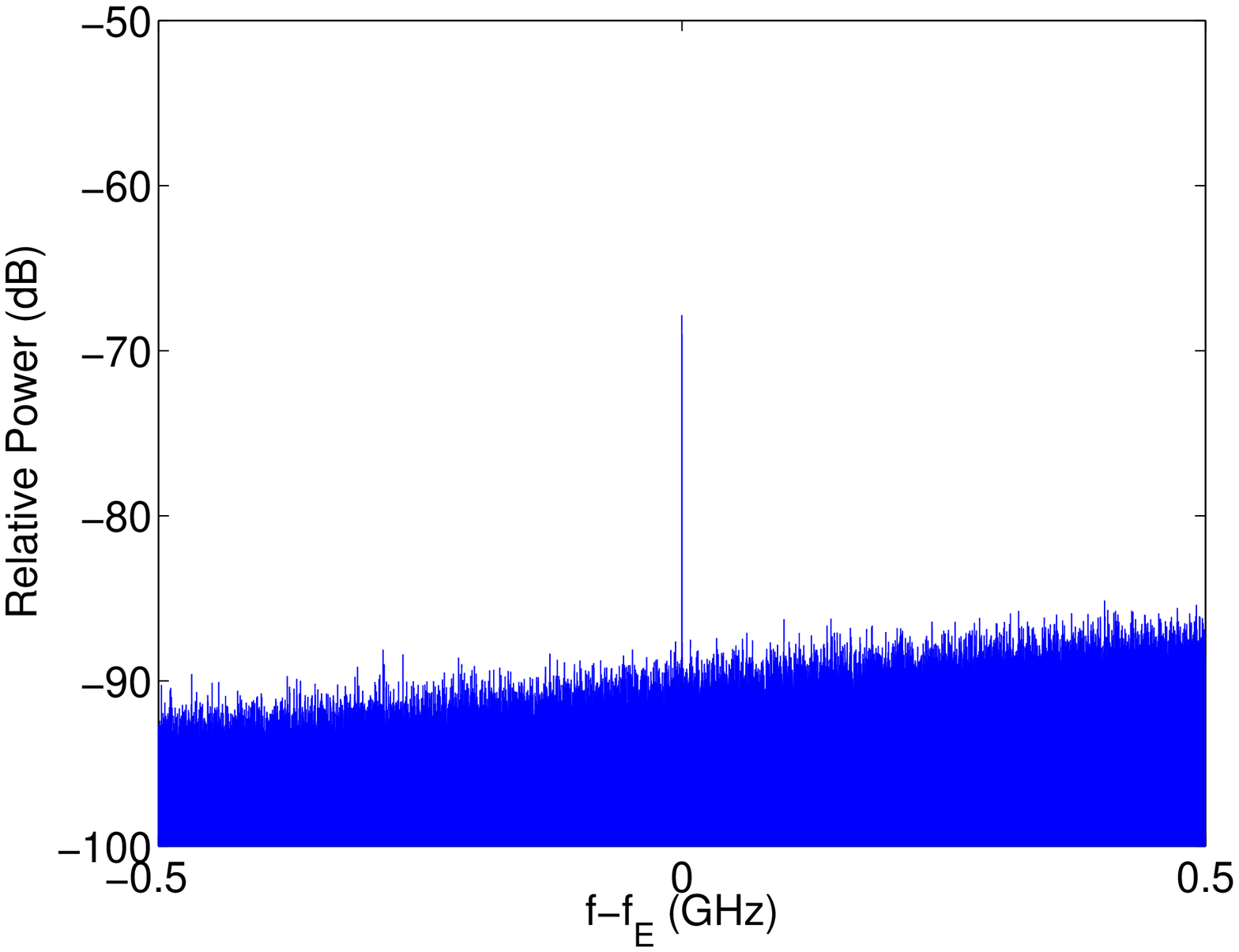}\\
{ (e)}&{ (f)}\\
\includegraphics[width=0.45\columnwidth]{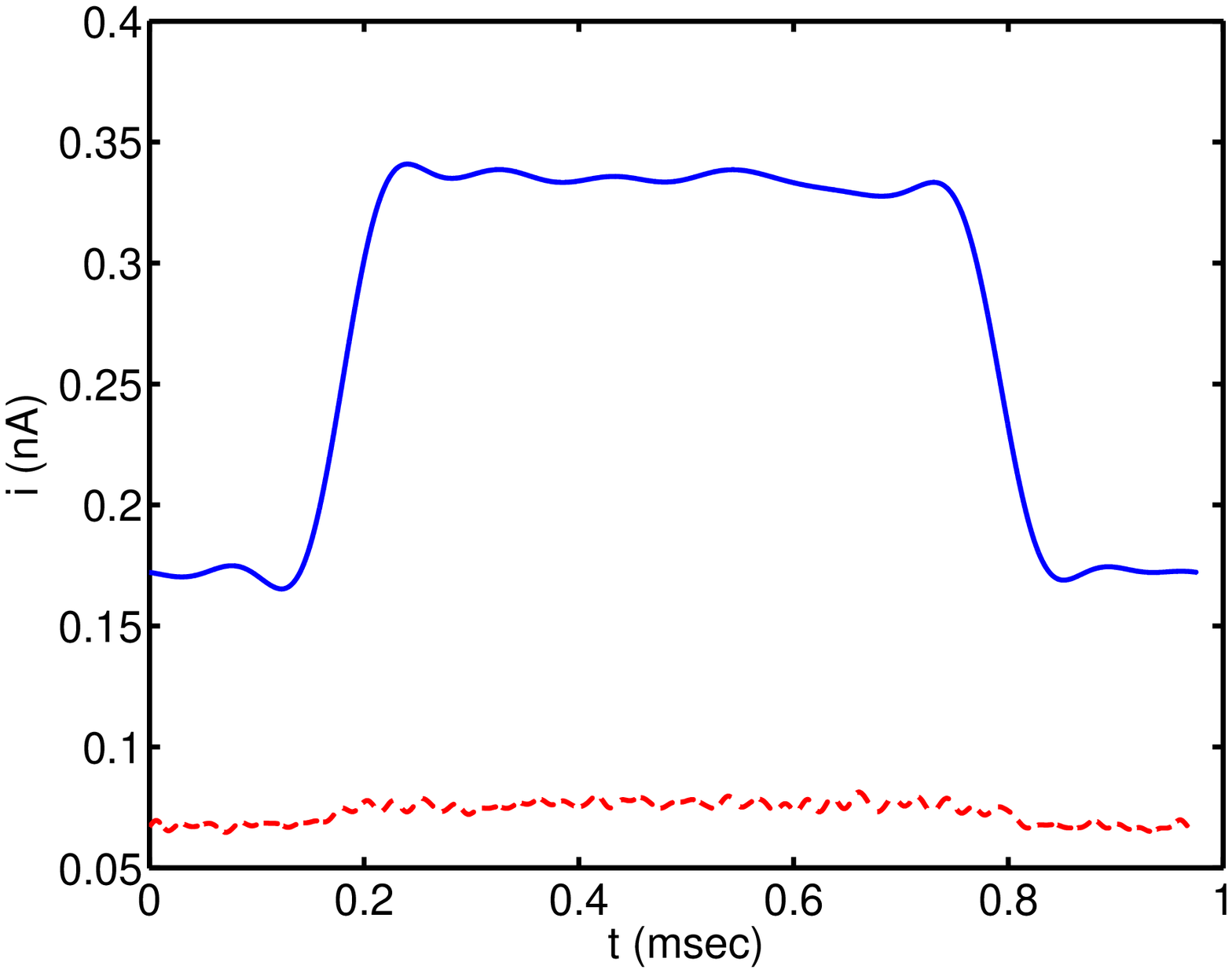}&\includegraphics[width=0.45\columnwidth]{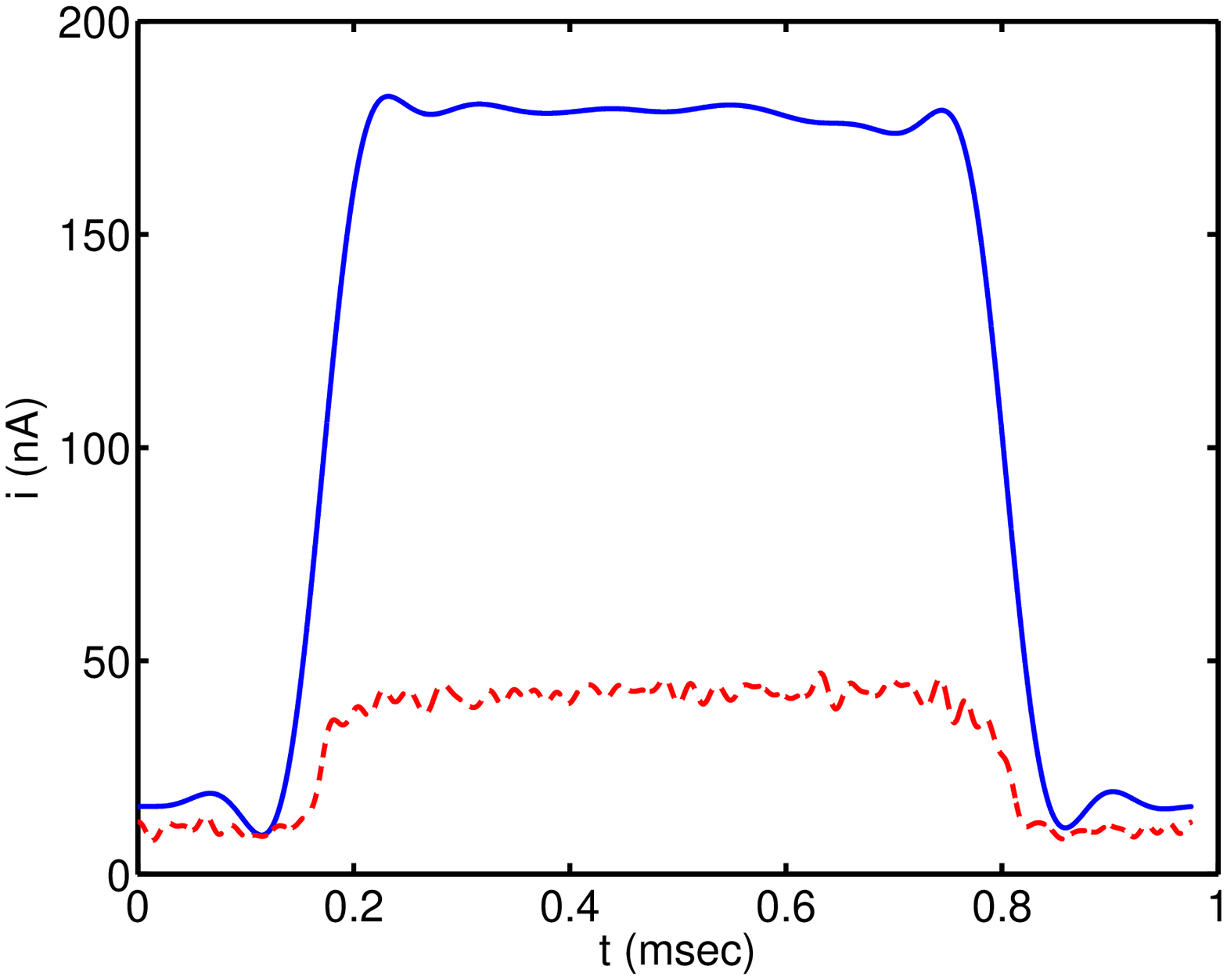}
\end{tabular}
\caption{Plots (a) and (b) display the optical spectra at Alice output coupler in the $01$ and $10$ states, respectively.  Marked in red is the contribution from Eve's signal.  (c) and (d) are the corresponding optical spectra at Eve's receiver after her phase modulator.  The fotocurrent measured by Eve is shown in (e) for a direct detection receiver and in (f) for a heterodyne coherent receiver.  Solid lines correspond to the 01 state and dashed to the 10 state.}\label{resultsBasic}
\end{figure}

Eve's laser is tuned to Alice's $0$ state grating and she modulates her carrier with a binary pseudorandom ${-1,+1}$ signal with a bin duration of $1$ ns.  The group delay of the grating is found to be negligible for this setup and the group delay associated with the propagation in a few meters of Erbium doped fiber has also been neglected.  In general, if the group delay at Alice's premises were relevant, Eve could always synchronize her transmitter and receiver phase modulators according to her knowledge of Alice's equipment.  The optical spectra seen at Alice and Eve's receivers and the photocurrent pulses measured by Eve are displayed in Fig. \ref{resultsBasic}.  Although Alice cannot detect the presence of Eve's signal,  Eve can perfectly distinguish the grating Alice is using since she has tuned her laser to the $0$ grating of Alice.  Both the direct detection and the coherent receiver would permit to determine the UFL state, even though the heterodyne receiver provides a significative improvement of the signal level.

\subsection{Length setup}

In this setup, Alice and Bob set their respective states by adding or not a section of $1$ km of fiber to the optical cavity instead of using gratings. We have assumed that both transmission ends include an optical filter with a Gaussian shape and $5$ GHz FWHM spectral width. All the EDFAs used both by Alice and Bob are assumed to have a small signal gain of $G=10$ dB, a saturation power of $Psat=13$ dBm, and a noise figure of $F=4.5$ dB. The optical power measured by Alice at his output coupler port during effective key transmission is $P_{out} \simeq 3.5$ mW.

Eve's laser frequency is again tuned $2.5$ GHz below the UFL center frequency and she modulates her carrier with a binary pseudorandom ${-1,+1}$ signal with a bin duration of $1$ ns.  Whereas a good synchronization exists between Eve's transmitted and received signal in the absence of the additional fiber segment, the presence of the additional path of $1$ km length makes the synchronization to be lost and permits Eve to determine Alice state.

\begin{figure}[!ht]
\centering

\begin{tabular}{cc}
{ (a)}&{ (b)}\\
\includegraphics[width=0.45\columnwidth]{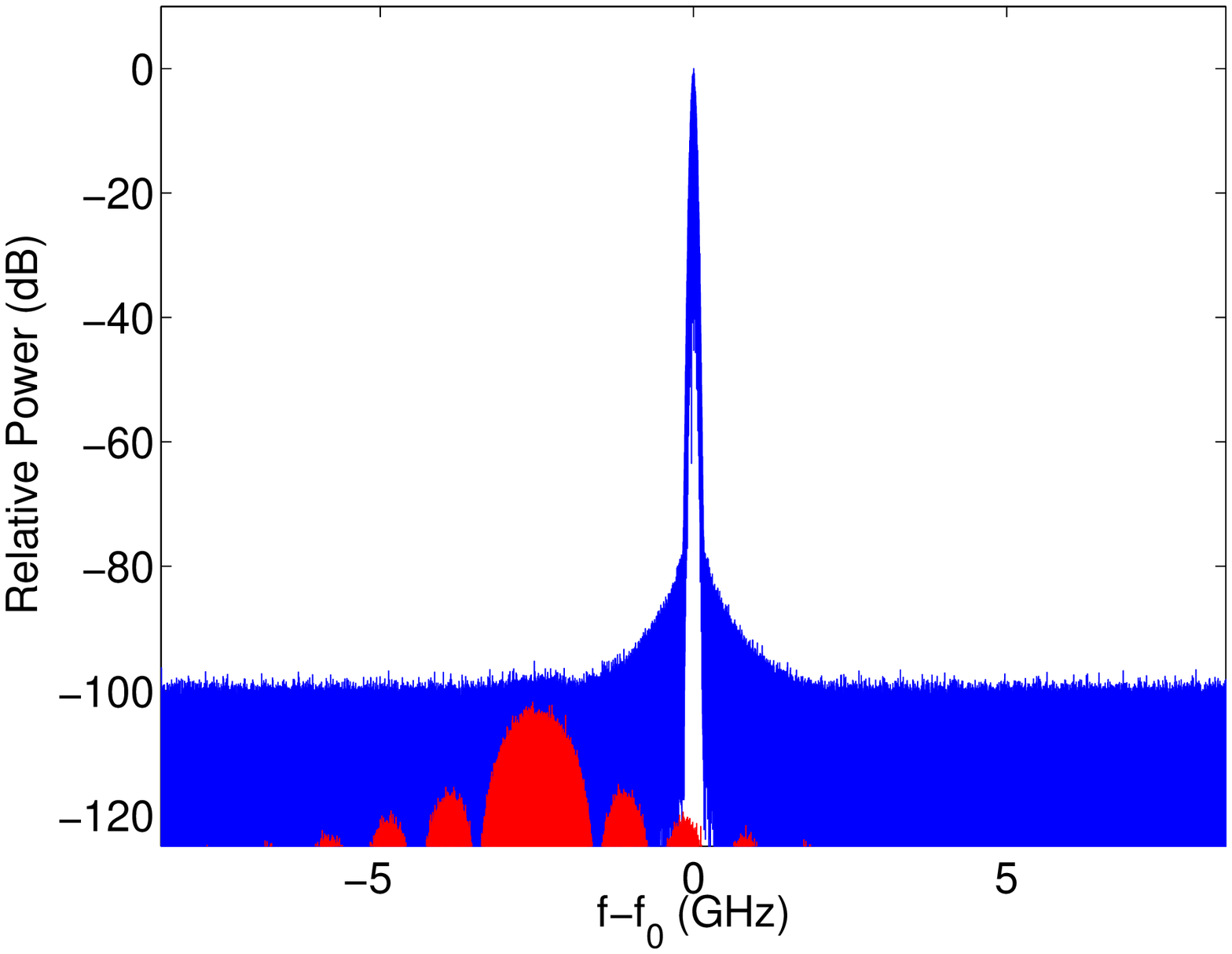}&\includegraphics[width=0.45\columnwidth]{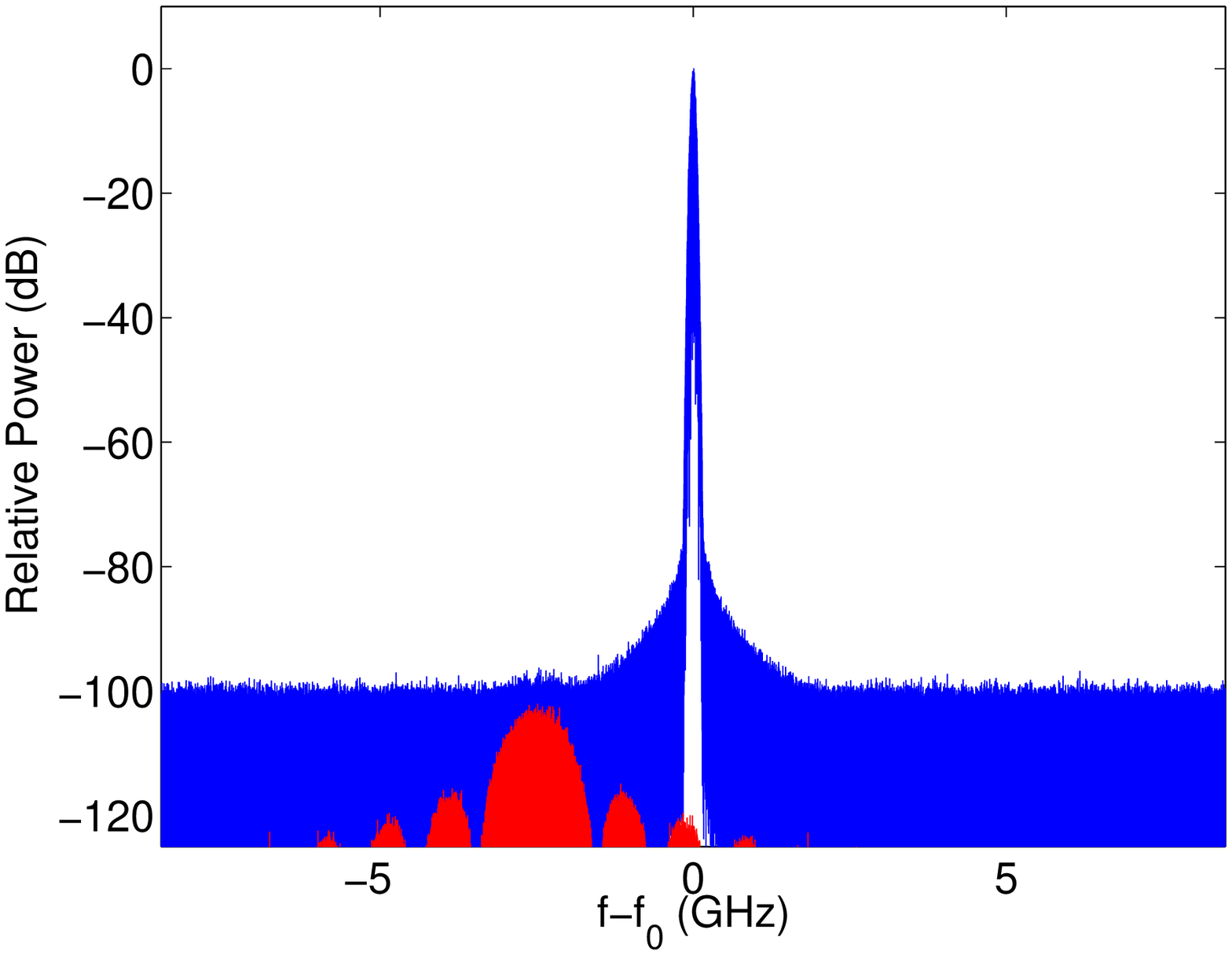}\\
{ (c)}&{ (d)}\\
\includegraphics[width=0.45\columnwidth]{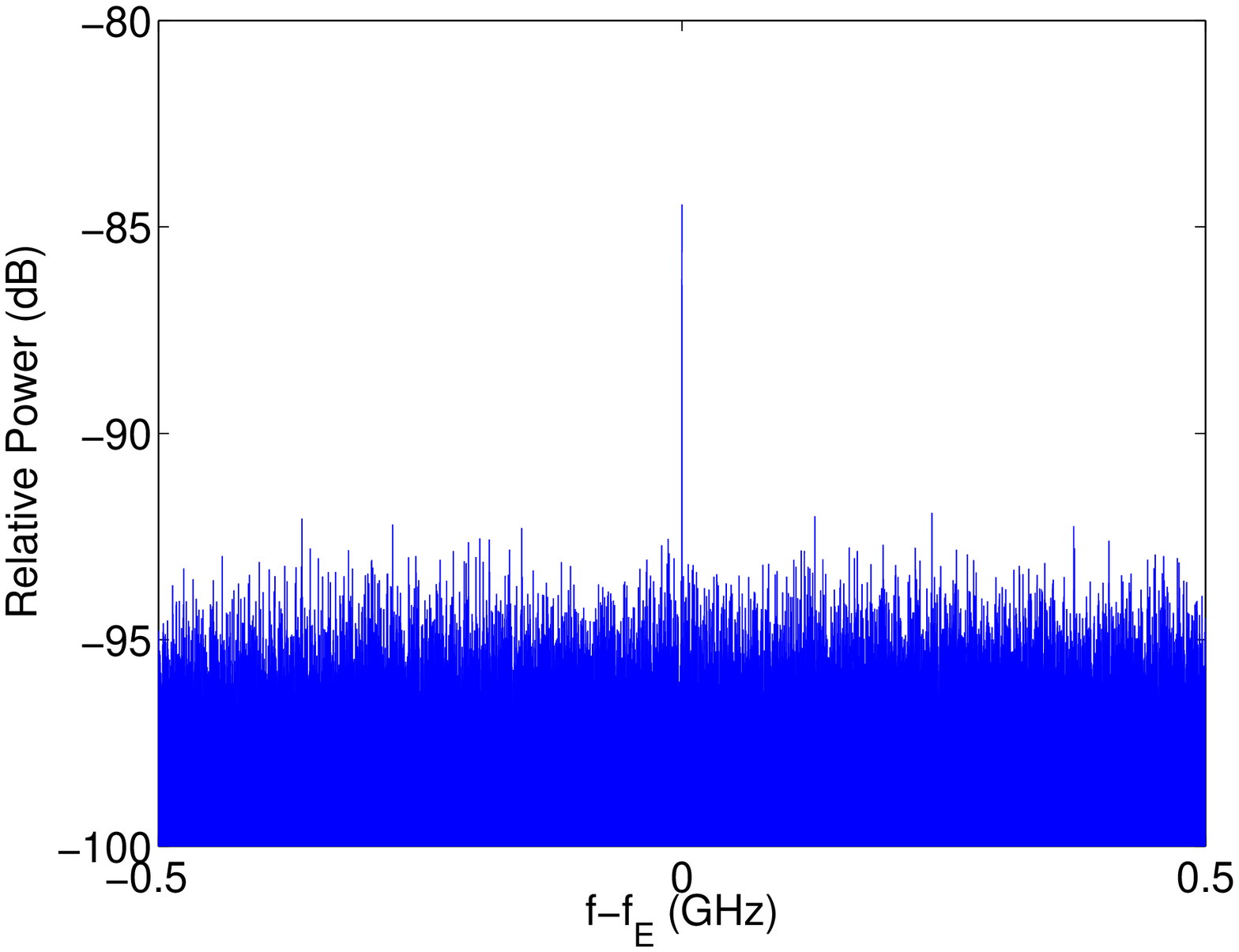}&\includegraphics[width=0.45\columnwidth]{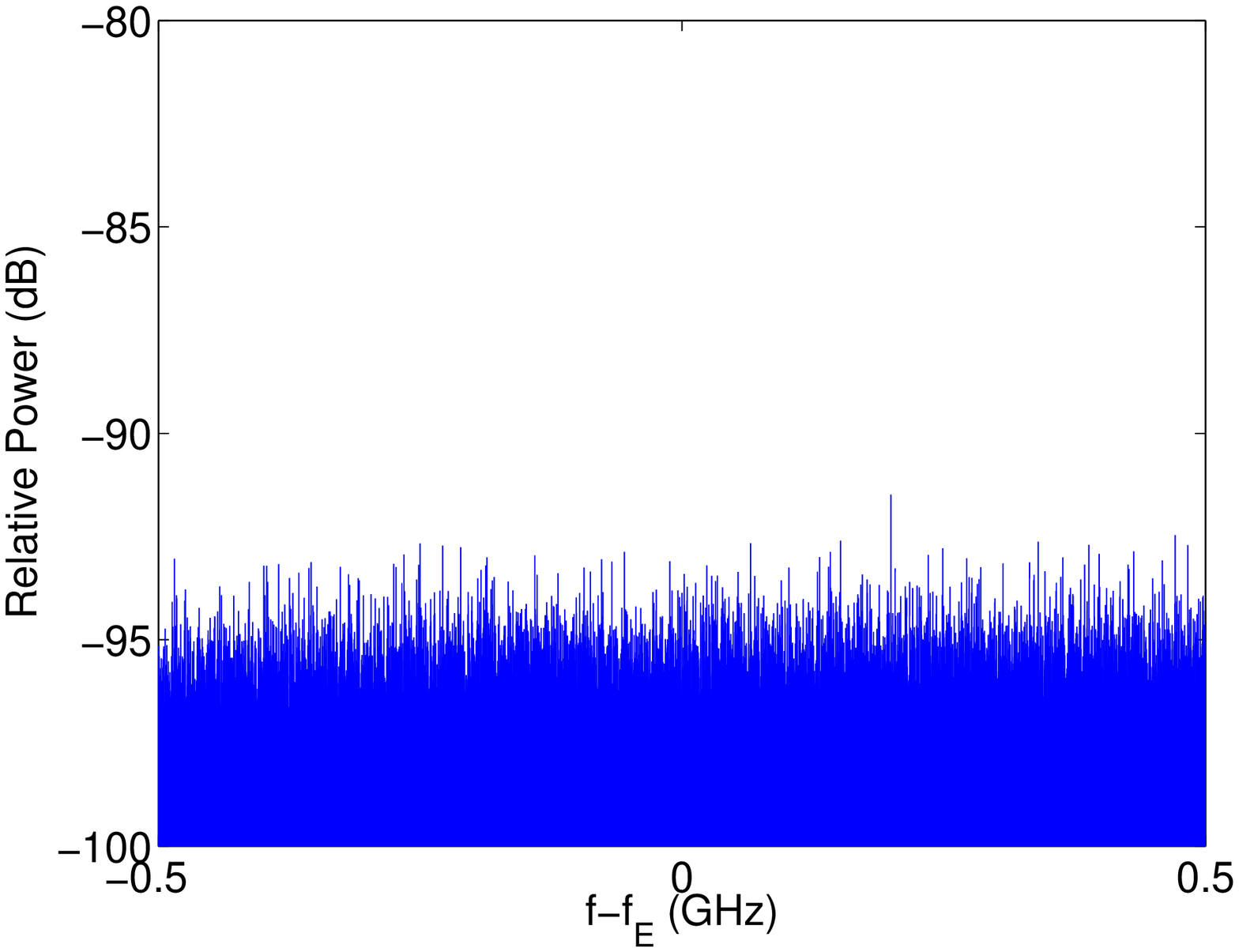}\\
{ (e)}&{ (f)}\\
\includegraphics[width=0.45\columnwidth]{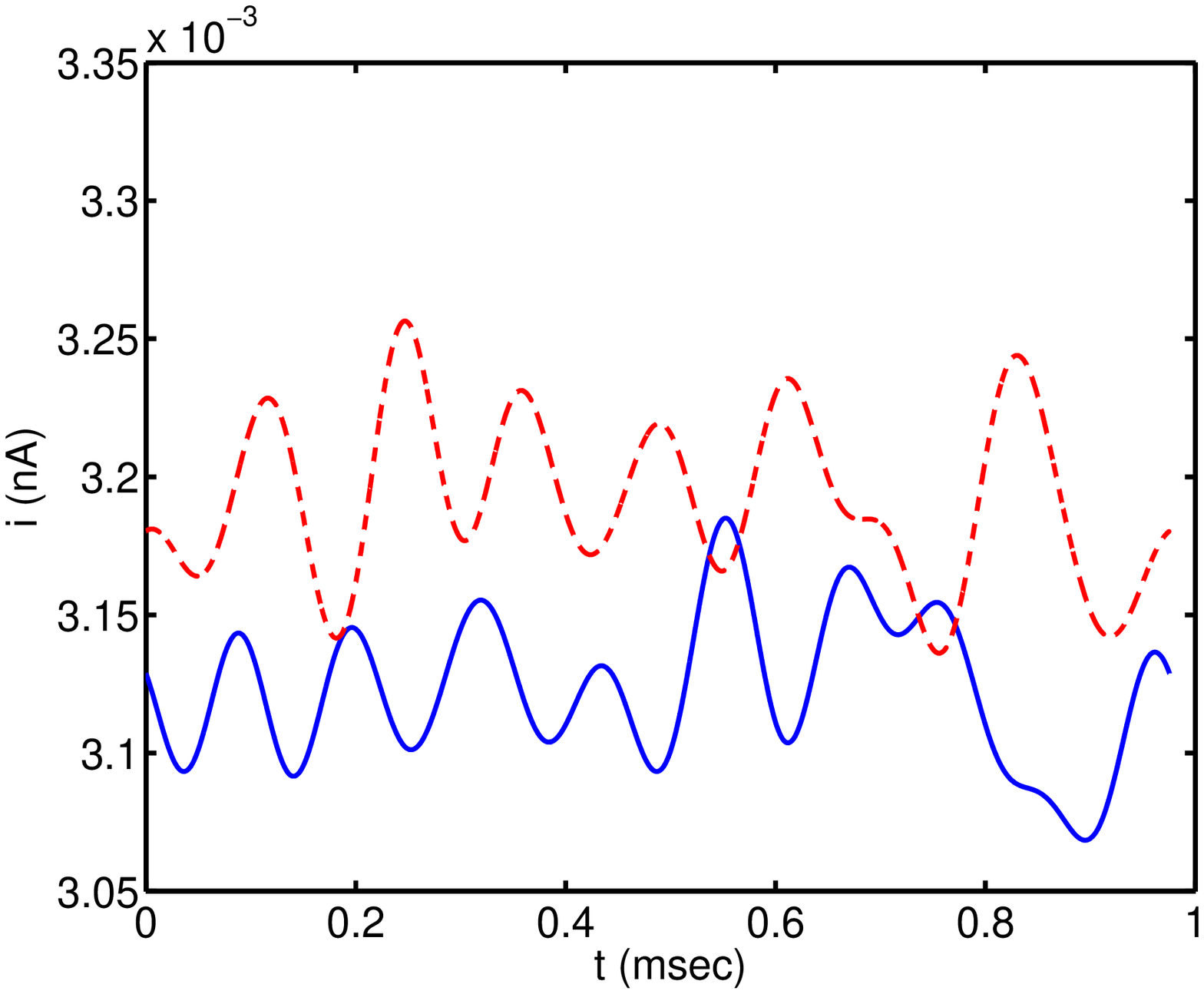}&\includegraphics[width=0.45\columnwidth]{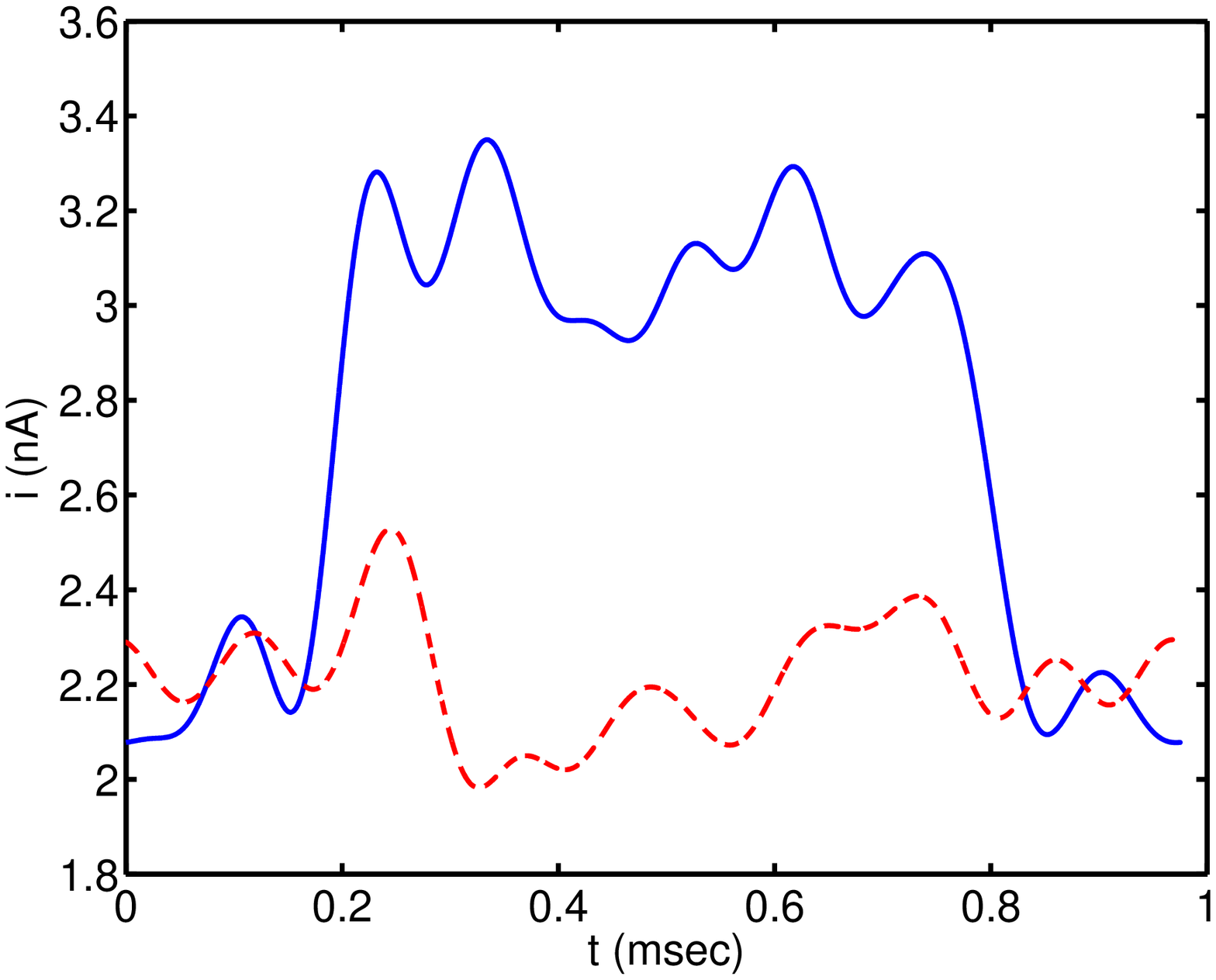}
\end{tabular}
\caption{Plots (a) and (b) display the optical spectra at Alice output coupler in the $01$ and $10$ states, respectively.  Marked in red is the contribution from Eve's signal.  (c) and (d) are the corresponding optical spectra at Eve's receiver after her phase modulator.  The fotocurrent measured by Eve is shown in (e) for a direct detection receiver and in (f) for a heterodyne coherent receiver.  Solid lines correspond to the 01 state and dashed to the 10 state.}\label{resultsLength}

\end{figure}

The optical spectra seen at Alice and Eve's receivers and the photocurrent pulses measured by Eve are displayed in Fig. \ref{resultsLength}.  Again, Alice cannot detect the presence of Eve's signal but Eve can perfectly distinguish the state of Alice.  In this case, only the heterodyne receiver permits to determine the UFL state.  As opposed to the previous case, now Eve's signal is out of the pass-band of the filter of Alice who listens to the channel before the Eve's signal has traversed the amplifiers.

\subsection{Dark states setup}
\label{DarkResults}

In this setup, all the UFL parameters are identical to those of the basic case but now Alice and Bob add a high-finesse Fabry-Perot optical filter with $\Delta f=25$ MHz FWHM bandwidth and $5$ GHz FSR at their respective transmission ends.  Secret key transmission now takes place at the non-lasing states of the UFL.  Even though the group delay of this filter is significative, Eve has to increase the bin duration of her phase modulator to $T_b=0.5$ $\mu$s to accommodate her probe signal into the narrow filter bandwidth.  For this modulation speed, the effect of the group delay is again negligible and her transmitter and receiver remain mutually synchronized.

\begin{figure}[!ht]
\centering

\begin{tabular}{cc}
{ (a)}&{ (b)}\\
\includegraphics[width=0.45\columnwidth]{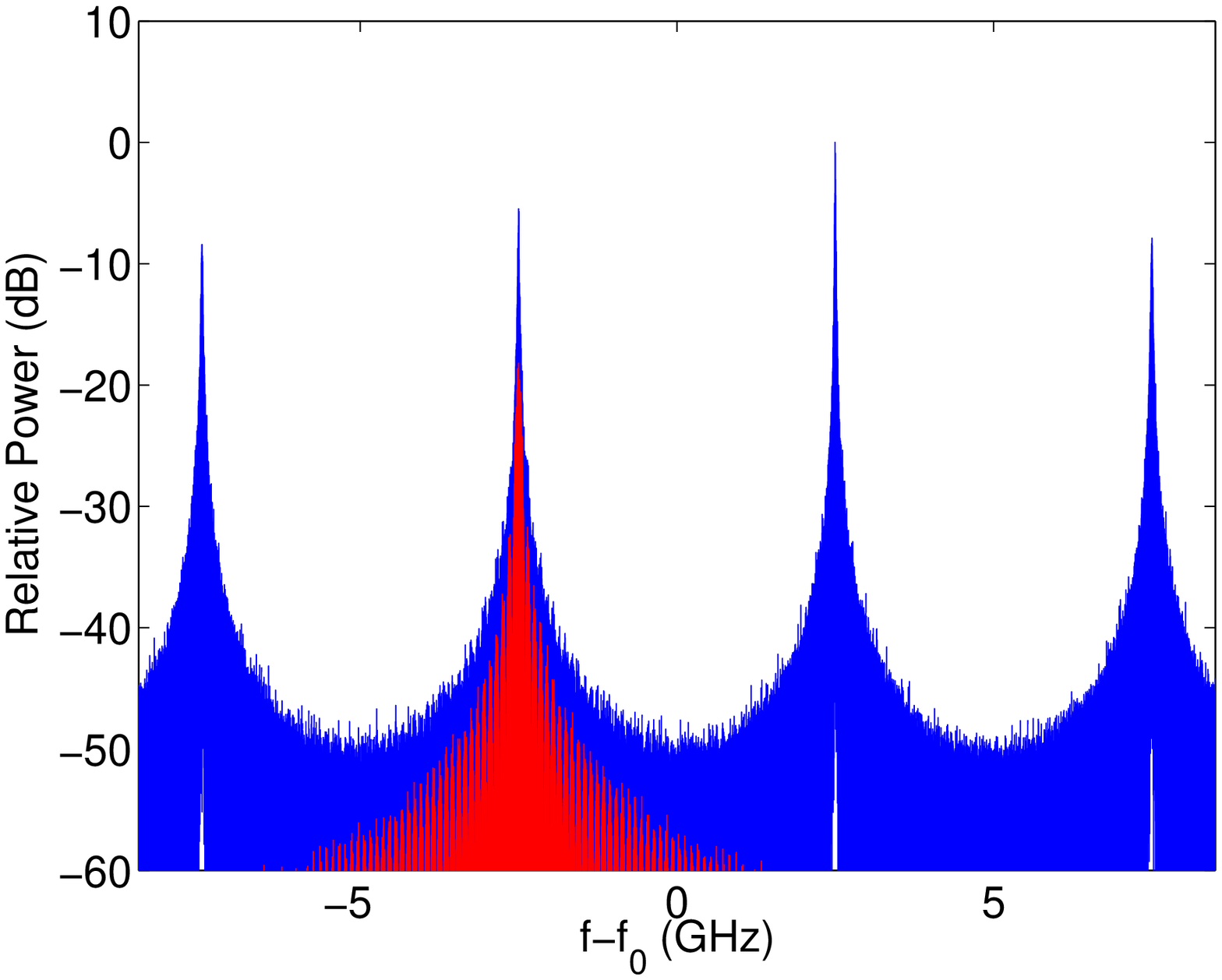}&\includegraphics[width=0.45\columnwidth]{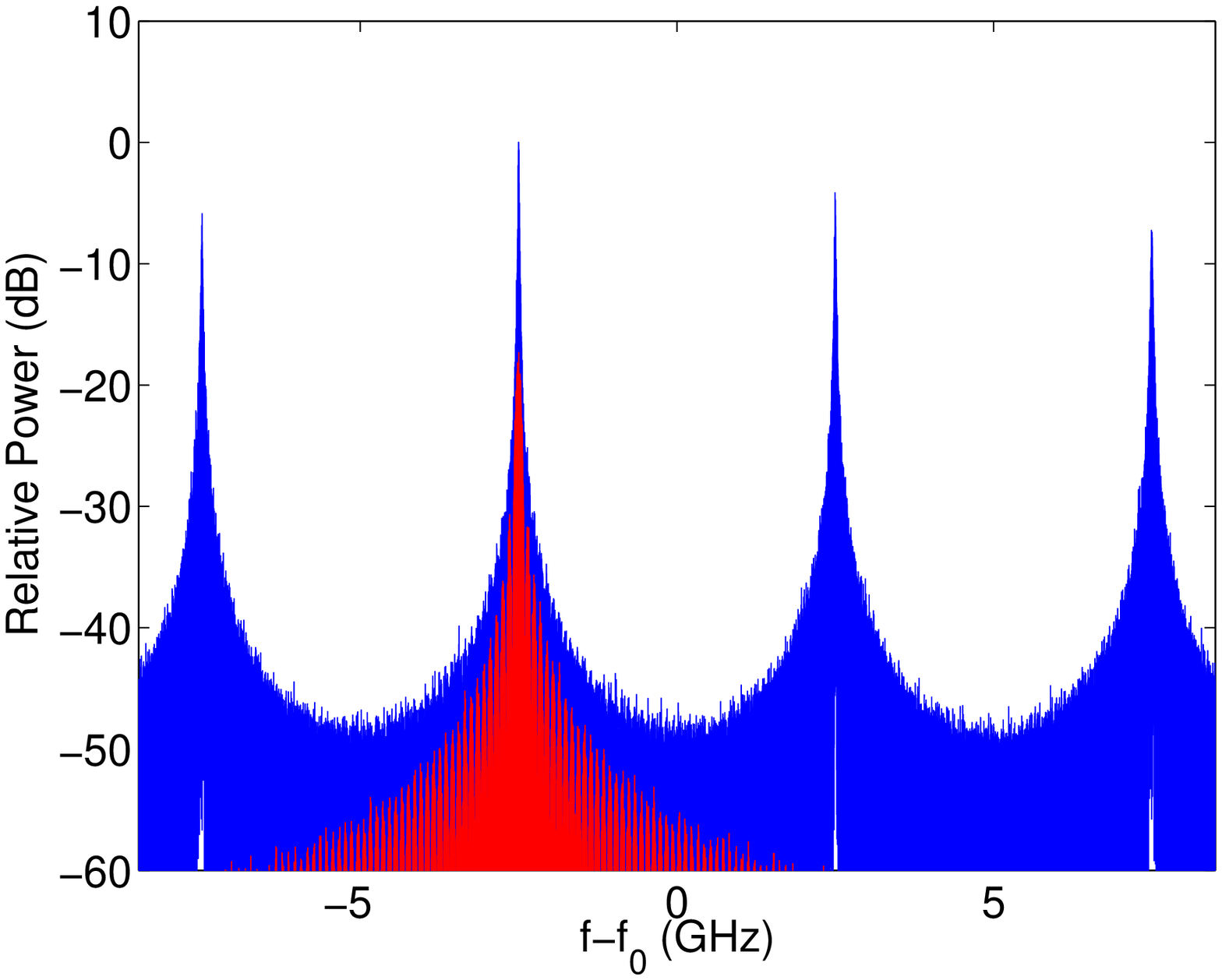}\\
{ (c)}&{ (d)}\\
\includegraphics[width=0.45\columnwidth]{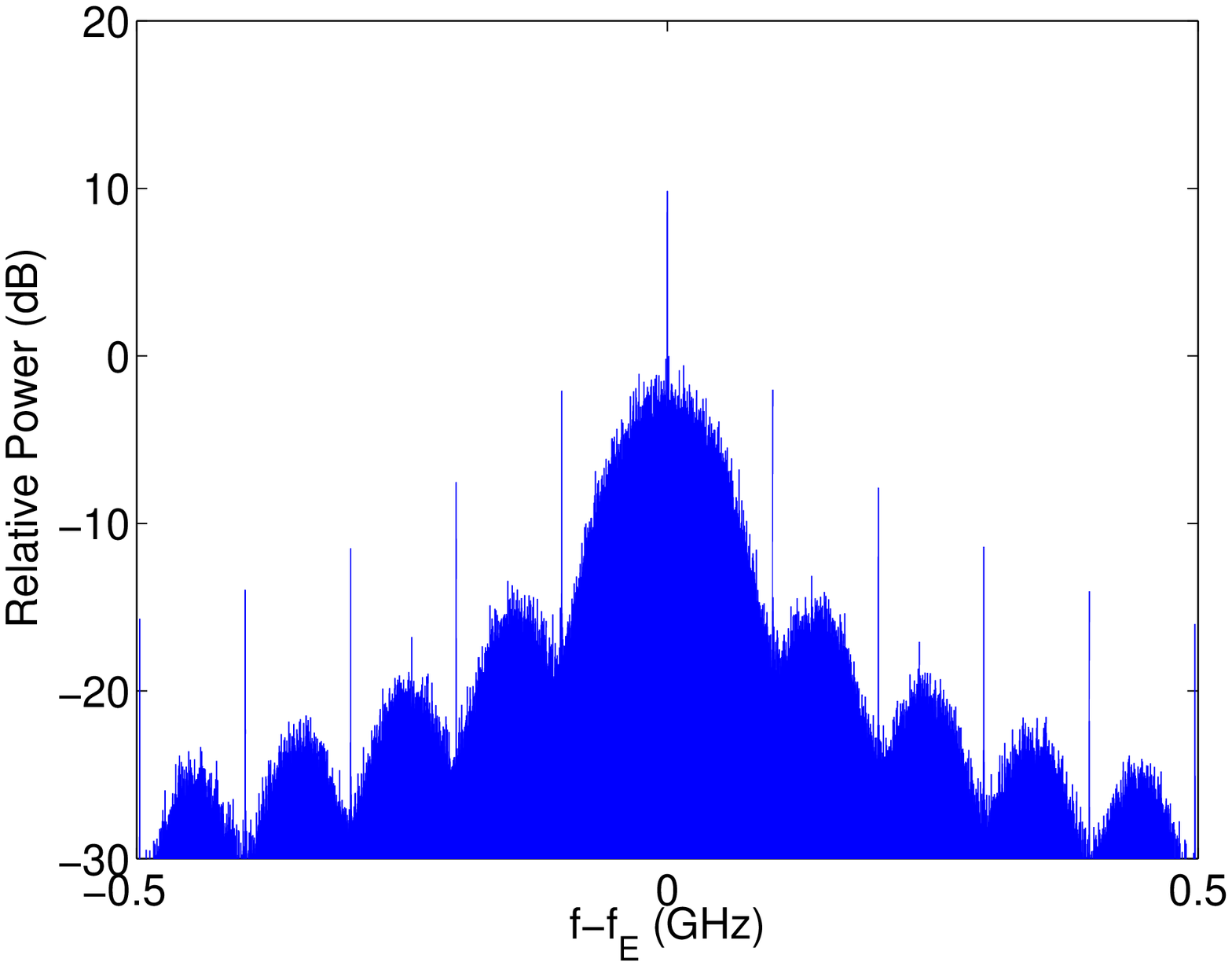}&\includegraphics[width=0.45\columnwidth]{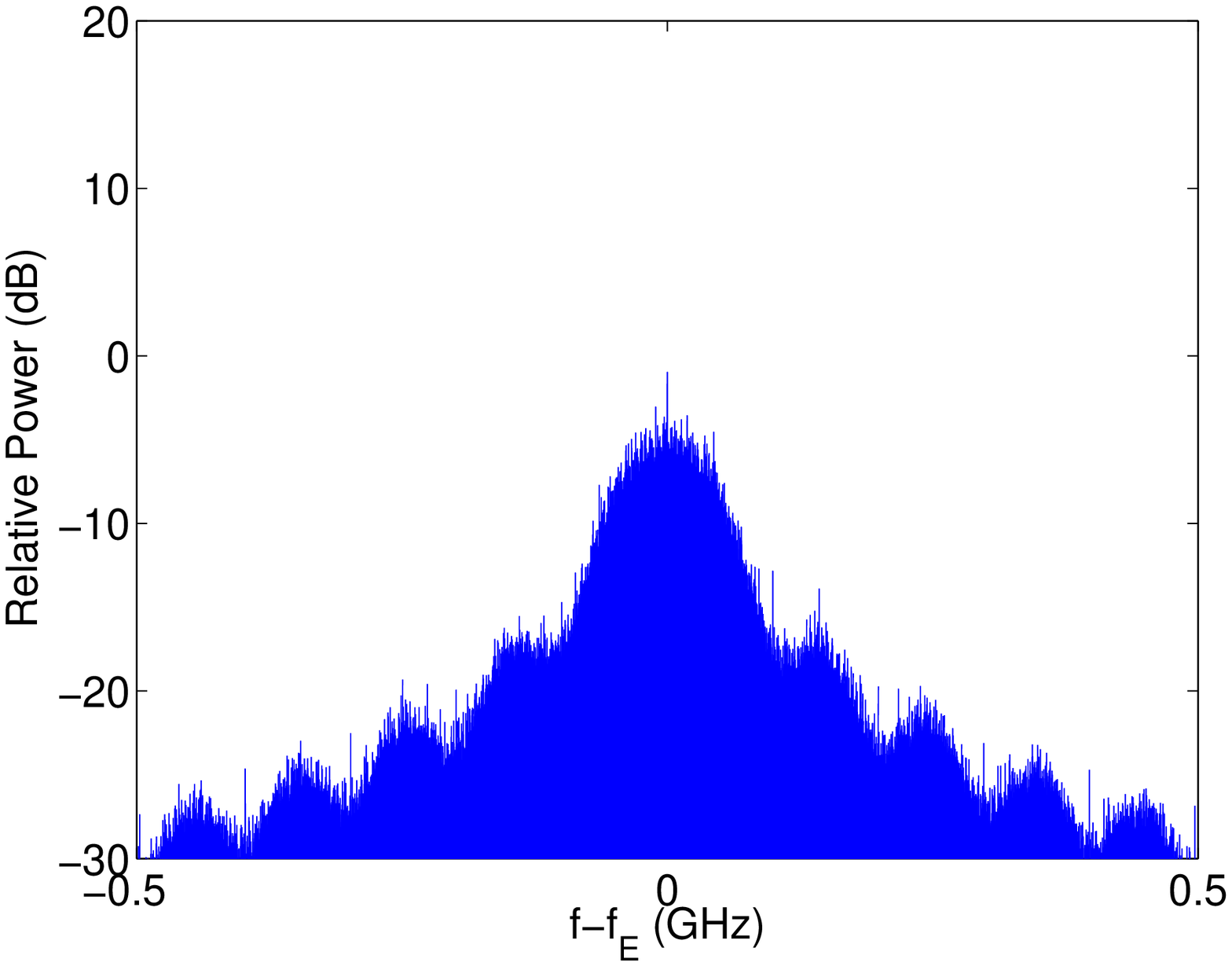}\\
{ (e)}&{ (f)}\\
\includegraphics[width=0.45\columnwidth]{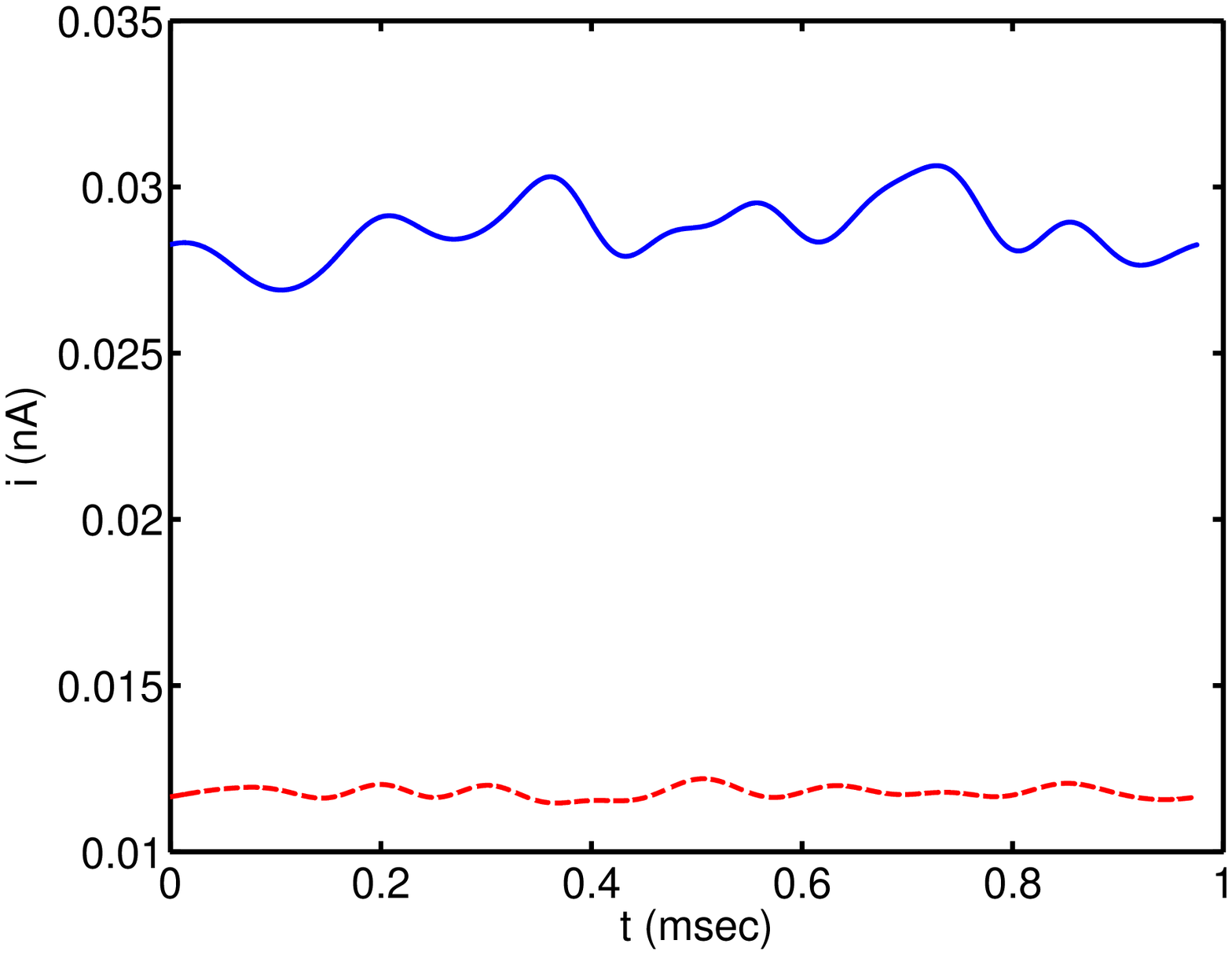}&\includegraphics[width=0.45\columnwidth]{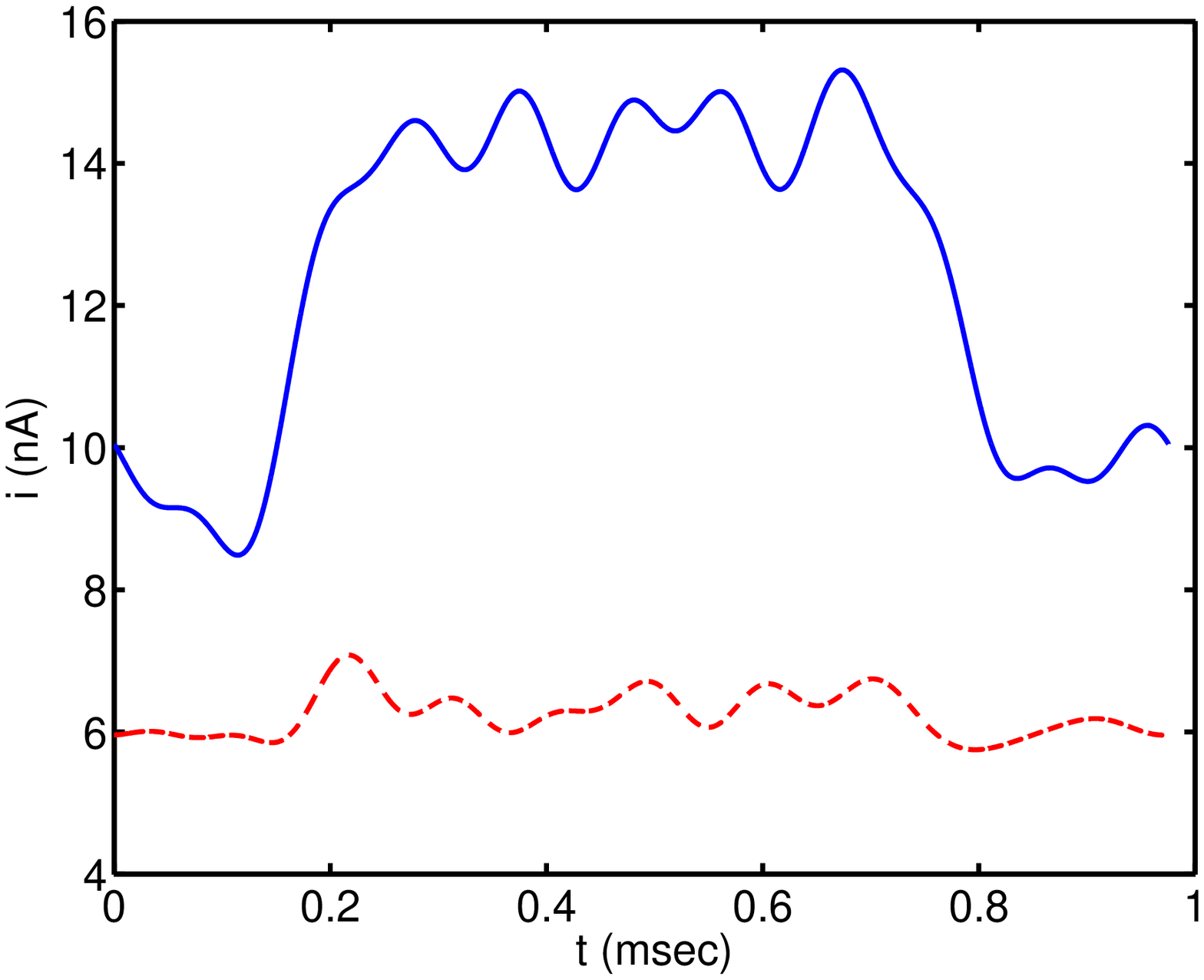}
\end{tabular}
\caption{Plots (a) and (b) display the optical spectra at Alice output coupler in the $01$ and $10$ states, respectively.  Marked in red is the contribution from Eve's signal.  (c) and (d) are the corresponding optical spectra at Eve's receiver after her phase modulator.  The fotocurrent measured by Eve is shown in (e) for a direct detection receiver and in (f) for a heterodyne coherent receiver.  Solid lines correspond to the 01 state and dashed to the 10 state.}\label{resultsDark}

\end{figure}

The optical spectra seen at Alice and Eve's receivers and the photocurrent pulses measured by Eve are displayed in Fig. \ref{resultsDark}.  Once again, Eve's signal is buried in noise at Alice receiver and she cannot detect the presence of Eve,  but Eve can distinguish the state of Alice since she has tuned her laser to the $0$ grating of Alice.  As in the length setup, only the heterodyne receiver permits the effective detection of Eve's signal.

 \section{Discussion: significance and countermeasures}
\label{Discussion}
We have presented a hidden probe attack where Eve can learn the configuration inside the end of an ultralong fiber laser key distribution system, allowing her to learn the established key. Using spread spectrum modulation, Eve can blend her probe inside the amplification noise in the channel and mask her presence. We have shown that she can attack different implementations of UFL key distribution with relatively simple equipment which depends on the concrete system under attack. The results show active attacks must be seriously considered in UFL key distribution. 

During simulation, we have found out that seemingly unimportant decisions, like where to put the coupler to sample the fiber's global state, can have important consquences. If Alice and Bob sample their signals before the gain stage, they can increase their chances of detecting any active attack. 

Additionally, proposed countermeasures to passive attacks, like intentionally using a noise source to mask the differences between the 01 and the 10 states \cite{SY06,ZSS08}, can backfire and make it easier to perform an active attack as it allows for a higher power in the probe pulses. Some other approaches against passive attacks, like inserting selective bandpass filters \cite{ZSS08}, also limit our active attack. However, in our simulations, the required bandwidths to effectively block the probe pulses also make the four state 00, 01, 10 and 11 difficult to distinguish and impose a burden on the legitimate users.

There are simple countermeasures that could reduce the threat posed by active hidden probe attacks. Our first suggestion is placing an attenuator before the gain stage but after the coupler used to sample the channel. Having a weaker signal before the amplifier will degrade the signal-to-noise ratio of the probe signal and Eve will need a higher optical power to achieve the same resolution. With the attenuator Alice and Bob increase their probability of detecting Eve's pulses with a minor disturbance to the global system. The important parameter for the fiber laser is the total gain. An additional loss can be compensated at the EDFA. Whether the attenuator is effective or not will depend on the level of noise, the bandwith and the time window for each bit of the key. If we could guarantee the SNR is so low that Shannon's noisy channel theorem gives a capacity of less than 1 bit per time window, that would be enough to discard undetected probes as a threat, but that seems out of reach for Alice and Bob with existing technology. However it points in the direction a resistant system should aim to force Eve to use advanced modulation and detection methods.

A second countermeasure would be introducing a random group delay inside Alice's and Bob's setups. For instance, introducing tension-controlled fiber Bragg gratings, Alice and Bob could introduce random group delays under the direction of a random sequence. Keeping the group delay unknown to Eve would not permit her to synchronize her phase modulators. This scheme could not be used in the length setup, though.

A third approach would be introducing a random phase filter to distort Eve's waveform. This approach has some limitations. The changes should be dynamic. Otherwise, Eve could probe the channel to learn its transfer function during the 00 and 11 states and use equalization to overcome any unexpected changes to the signal. Microring devices could give the desired effect and introduce a random phase that would distort the probe's waveform \cite{BHV12}. 

Any succesful solution will need to balance the complexity on the legitimate users and give, at the same time, enough protection against active attacks. An important open question is whether some of the methods of optical steganography \cite{WPN06,WN06,HWG07,FP09,WFX10,WP11,WWT13,WWS14,WSM15} could improve the abilities of Eve and whether there are or not effective countermeasures against the most general kind of active attacks that try to learn the chosen configuration at the fiber ends by sending exploration pulses and observing their changes after crossing the system of Alice or Bob.

\section*{Acknowledgment}
This research has been funded by Project TEC2015-69665-R (MINECO/FEDER, UE) and Junta de Castilla y Le\'on Project No. VA089U16.

\ifCLASSOPTIONcaptionsoff
  \newpage
\fi



\bibliographystyle{IEEEtran}
%
\newcommand{\noopsort}[1]{} \newcommand{\printfirst}[2]{#1}
  \newcommand{\singleletter}[1]{#1} \newcommand{\switchargs}[2]{#2#1}

%

\begin{IEEEbiography}[{\includegraphics[width=1in,height=1.25in,clip,keepaspectratio]{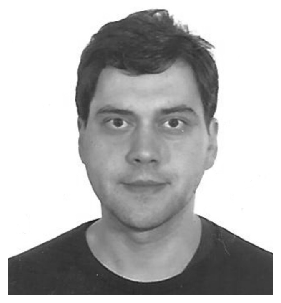}}]{Juan Carlos Garc\'ia-Escart\'in}
J.C. Garc\'ia-Escart\'in obtained his degree in Telecommunication Engineering from the Universidad de Valladolid, Spain, with a master thesis in the Norwegian University from Science and Technology (NTNU) in Trondheim, Norway. He received his Ph.D. in 2008 from the Universidad de Valladolid, Spain, where he is currently teaching. His interests include quantum information, quantum optics and the relationship between physics and information.
\end{IEEEbiography}

\begin{IEEEbiography}[{\includegraphics[width=1in,height=1.25in,clip,keepaspectratio]{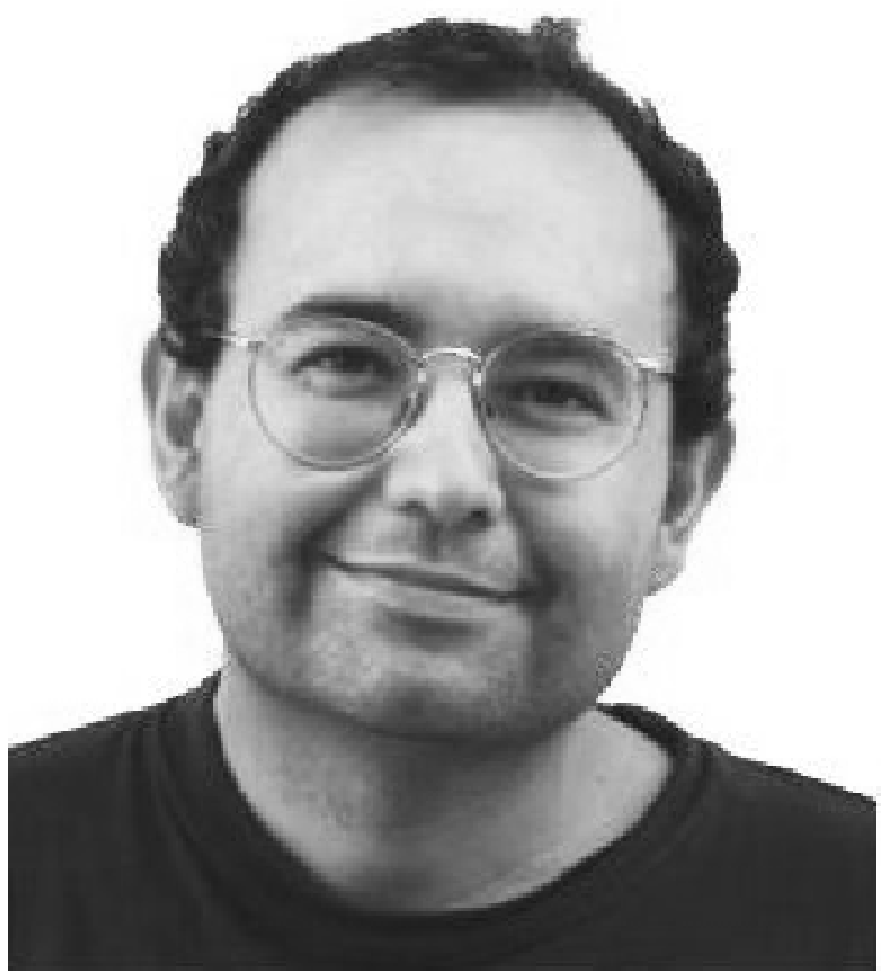}}]{Pedro Chamorro-Posada}
P. Chamorro-Posada received the M.S. and Ph.D. degrees in Telecommunication Engineering from the Universidad de Vigo, Spain, in 1992 and 1995, respectively. He is currently a Professor at the Universidad de Valladolid, Spain. His research areas include quantum information, optical solitons, photonic devices and optical materials.
\end{IEEEbiography}





\end{document}